\documentclass[a4paper,number,5p,twocolumn]{elsarticle}

\usepackage[english]{babel}
\usepackage[utf8]{inputenc}
\usepackage{graphicx}
\usepackage[usenames,dvipsnames]{xcolor}
\usepackage{bm}
\usepackage{amsfonts}
\usepackage{amssymb}
\usepackage{amsmath}	
\usepackage{epstopdf}
\usepackage{soul}
\usepackage[normalem]{ulem}
\usepackage[switch]{lineno}
\usepackage{setspace}  
\usepackage{csquotes}
\usepackage{algorithm}
\usepackage{algpseudocode}
\usepackage[	
    unicode,
    breaklinks=true,
    hypertexnames=false,
    colorlinks=false,
    citecolor=black,
    filecolor=black,
    linkcolor=black,
    urlcolor=black
]{hyperref}
\usepackage{natbib}
\usepackage{mathtools}
\usepackage{interval}
\usepackage{multirow}
\usepackage{array}

\newcommand{\rem}[1]{}

\newcommand{\quest}[1]{#1}
\newcommand{\replace}[2]{\rem{#1}{\quest{#2}}}

\newcommand{\Fref}[1]{Fig.~\ref{#1}}
\newcommand{\Tref}[1]{Tab.~\ref{#1}}	
\newcommand{\Eref}[1]{Eq.~(\ref{#1})}

\newcommand{\Sref}[1]{{Section~\ref{#1}}}

\newcommand{\etal}{~et~al.}

\newcommand{\homg}{\textrm{hom}}

\newcommand{\avgsdom}[2]{\left\langle #1 \right\rangle_{#2}} 	
\newcommand{\Stwo}{\ensuremath{S_2}}

\newcommand{\temp}{\theta}
\newcommand{\tempgrad}{g}
\newcommand{\Tempgrad}{G}
\newcommand{\flux}{q}

\newcommand{\K}{K}

\newcommand{\Kapp}{\tenss{\K}_{\svesize}^{\mathcal{B}}}
\newcommand{\Khom}{\tenss{\K}^{\homg}}

\newcommand{\Strain}{E}
\newcommand{\strain}{\varepsilon}

\newcommand{\displacement}{u}

\newcommand{\fdispl}{\tens{\displacement}^{\ast}}
\newcommand{\vardispl}{\tilde{\tens{\displacement}}}
\newcommand{\varfdispl}{\tilde{\tens{\displacement}}^{\ast}}

\newcommand{\stress}{\sigma}

\newcommand{\D}{D}
\renewcommand{\C}{C}
\newcommand{\Dtens}{\tensf{\D}}
\newcommand{\Ctens}{\tensf{\C}}
\newcommand{\Dhom}{\tensf{\D}^{\homg}}
\newcommand{\Dapp}{\tensf{\D}_{\svesize}^{\mathcal{B}}}

\newcommand{\bmath}[1]{\ensuremath{\bm{#1}}}

\mathchardef\mhyphen="2D

\newcommand{\de}[1]{\,{\mathrm d}#1}

\newcommand{\atx}{\ensuremath{(\tens{x})}}

\newcommand{\set}[1]{{\mathbb #1}}
\newcommand{\setN}{\set{N}}
\newcommand{\setR}{\set{R}}
\newcommand{\setRn}[1]{\set{R}^{#1}}

\newcommand{\domain}{\Omega}
\newcommand{\domainsize}{\Omega_{\svesize}}

\newcommand{\boundarysize}{\partial\domainsize}

\newcommand{\norm}[1]{\left\lVert#1\right\rVert}
\newcommand{\measure}[1]{|#1|}

\newcommand{\scal}[1]{\mathnormal{#1}}
\newcommand{\tens}[1]{\boldsymbol{#1}}					
\newcommand{\tenss}[1]{\bmath{#1}} 					    
\newcommand{\tensf}[1]{\bmath{\mathbf{#1}}} 			

\newcommand{\x}{\tens{x}} 								

\newcommand{\scontr}{\cdot}
\newcommand{\dcontr}{\,\colon}

\newcommand{\svesize}{\ensuremath{s}}
\newcommand{\svesizemax}{\ensuremath{s_{\textrm{max}}}}
\newcommand{\real}{\ensuremath{n_\svesize}}
\newcommand{\realmin}{\ensuremath{n_{\textrm{min}}}}
\newcommand{\realmax}{\ensuremath{n_{\textrm{max}}}}

\newcommand{\charlen}{\ensuremath{\chi}}
\newcommand{\volfrac}{\ensuremath{\phi}}
\newcommand{\contrast}{\ensuremath{\kappa}}
\newcommand{\RVEsizeCharLen}{\ensuremath{\varrho}}

\newcommand{\bias}{\delta}
\newcommand{\std}{\varsigma}
\newcommand{\sampleAver}[1]{\ensuremath{\hat{{#1}}_{\svesize}}}
\newcommand{\sampleStd}[1]{\ensuremath{\hat{\std}_{\svesize}^{#1}}}
\newcommand{\scalOne}[1]{\ensuremath{{#1}_{\svesize}^{(i)}}}
\newcommand{\scalMean}[1]{\ensuremath{\overline{{#1}_{\svesize}}}}

\newcommand{\Lletter}{l}
\newcommand{\Ltens}{\ensuremath{\tenss{\MakeUppercase{\Lletter}}}}
\newcommand{\LtensOne}{\Ltens^{(i)}_{\svesize}}
\newcommand{\LtensHom}{\ensuremath{\Ltens^{\homg}}}
\newcommand{\Lscal}{\ensuremath{\scal{\MakeLowercase{\Lletter}}}}
\newcommand{\LscalOne}{\scalOne{\Lscal}}
\newcommand{\LscalHom}{\ensuremath{\Lscal^\homg}}
\newcommand{\LscalBias}{\ensuremath{\bias^{\Lscal}_{\svesize}}}
\newcommand{\LscalError}{\ensuremath{\scal{e}^{\Lscal,(i)}_{\svesize}}}
\newcommand{\LscalMean}{\ensuremath{\scalMean{\Lscal}}}
\newcommand{\LscalErrorStd}{\ensuremath{\std_{\svesize}^\Lscal}}

\newcommand{\Mletter}{m}
\newcommand{\Mtens}{\ensuremath{\tenss{\MakeUppercase{\Mletter}}}}
\newcommand{\MtensOne}{\Mtens^{(i)}_{\svesize}}
\newcommand{\MtensHom}{\ensuremath{\Mtens^{\homg}}}
\newcommand{\Mscal}{\ensuremath{\scal{\MakeLowercase{\Mletter}}}}
\newcommand{\MscalOne}{\ensuremath{\Mscal_\svesize^{(i)}}}
\newcommand{\MscalHom}{\ensuremath{\Mscal^\homg}}
\newcommand{\MscalBias}{\ensuremath{\bias^{\Mscal}_{\svesize}}}
\newcommand{\MscalError}{\ensuremath{\scal{e}^{\Mscal,(i)}_{\svesize}}}
\newcommand{\MscalMean}{\ensuremath{\scalMean{\Mscal}}}
\newcommand{\MscalErrorStd}{\ensuremath{\std_{\svesize}^\Mscal}}

\newcommand{\errorSize}{\epsilon}
\newcommand{\errorSizeLim}{\epsilon^{\textrm{usr}}}
\newcommand{\errorProximity}{\xi}
\newcommand{\errorProximityOne}{\scalOne{\errorProximity}}
\newcommand{\errorProximityLim}{\errorProximity^{\textrm{usr}}}
\newcommand{\errorProximityTest}{\errorProximity_{\svesize}^{\textrm{test}}}

\newcommand{\tdist}[2]{t_{{#2}}^{-1}{({\textstyle{#1}})}\,}
\newcommand{\confLevelOne}{\tilde{P}_{\errorSize}}
\newcommand{\confLevelTwo}{\tilde{P}_{\errorProximity}}

\setstcolor{BrickRed}

\newcommand{\oneColFigWidth}{0.8\columnwidth}
\newcommand{\halfColFigWidth}{0.495\columnwidth}
\newcommand{\rempunct}[1]{}

\DeclareGraphicsExtensions{.pdf,.eps,.png,.jpg}

\soulregister\cite7
\soulregister\ref7
\soulregister\pageref7

\makeatletter
\providecommand{\doi}[1]{%
	\begingroup
	\let\bibinfo\@secondoftwo
	\urlstyle{rm}%
	\href{http://dx.doi.org/#1}{%
		doi:\discretionary{}{}{}%
		\nolinkurl{#1}%
	}%
	\endgroup
}
\makeatother

\makeatletter
\def\ps@pprintTitle{
	\let\@oddhead\@empty
	\let\@evenhead\@empty
	\def\@oddfoot{{\small© 2018. This manuscript version is made available under the \href{http://creativecommons.org/licenses/by-nc-nd/4.0/}{ CC-BY-NC-ND 4.0 license}.}\hfill{}}%
	\let\@evenfoot\@oddfoot
}
\makeatother

\begin{document}

\begin{frontmatter}

\title{Wang tiling aided statistical determination of the Representative Volume Element size of random heterogeneous materials\tnoteref{t1}}
\tnotetext[t1]{Author's post-print version of the article manuscript published in \mbox{\textit{European Journal of Mechanics -- A/Solids}}\\\href{http://dx.doi.org/10.1016/j.euromechsol.2017.12.002}{DOI: 10.1016/j.euromechsol.2017.12.002}.}

\author[ctu]{Martin Do\v{s}k\'{a}\v{r}\corref{cor}}
\ead{martin.doskar@fsv.cvut.cz}
\author[ctu,utia]{Jan Zeman}
%
\author[ctu]{Daniela Jaru\v{s}kov\'{a}}
%
\author[ctu]{Jan Nov\'{a}k}
%
\address[ctu]{Faculty of Civil Engineering, Czech Technical University in Prague, Th\'{a}kurova 7, \mbox{166 29 Prague 6}, Czech Republic}
\address[utia]{Institute of Information Theory and Automation, Academy of Sciences of the Czech Republic, Pod Vodárenskou věží 4, \mbox{182 08 Prague 8}, Czech Republic}

\cortext[cor]{Corresponding author.}

\journal{arXiv.org}

\begin{abstract}
	
Wang tile based representation of a heterogeneous material facilitates fast synthesis of non-periodic microstructure realizations. In this paper, we apply the tiling approach in numerical homogenization to determine the Representative Volume Element size related to the user-defined significance level and the discrepancy between bounds on the apparent properties. First, the tiling concept is employed to efficiently generate arbitrarily large, statistically consistent realizations of investigated microstructures. Second, benefiting from the regular structure inherent to the tiling concept, the Partition theorem, and statistical sampling, we construct confidence intervals of the apparent properties related to the size of a microstructure specimen. Based on the interval width and the upper and lower bounds on the apparent properties, we adaptively generate additional microstructure realizations in order to arrive at an RVE satisfying the prescribed tolerance. The methodology is illustrated with the homogenization of thermo-mechanical properties of three two-dimensional microstructure models: a microstructure with mono-disperse elliptic inclusions, foam, and sandstone.

\end{abstract}

\begin{keyword}
	Representative Volume Element size; Wang tiling; numerical homogenization
\end{keyword}

\end{frontmatter}

\section{Introduction}
\label{sec:introduction}

The Representative Volume Element (RVE) is the key concept in modelling of heterogeneous materials. The original definition by Hill~\cite{hill_elastic_1963} requires an RVE to (i) be \enquote{structurally entirely typical of the whole mixture on average} and (ii) \enquote{contain a sufficient number of inclusions for the apparent overall moduli to be effectively independent of the surface values of traction and displacement, so long as these values are \enquote{macroscopically uniform}.} For materials with periodic microstructure, these requirements are met by any periodic part of the microstructure under periodic boundary conditions~\cite{ostoja-starzewski_material_2006}.

However, the majority of real-world materials display randomness in their microstructures. \citet{sab_homogenization_1992} proved that microstructure ergodicity and statistical homogeneity are the essential requirements for the existence of an RVE. 
He also showed that the second Hill requirement is attainable only in the infinite-size limit and, thus, homogenized properties determined from any finite-size microstructure realization are biased by the adopted boundary conditions. For this reason, an error measure and its threshold have to be introduced in order to define an RVE for random heterogeneous materials (also referred to as a \enquote{computational RVE}~\cite{salmi_various_2012} or a \enquote{numerical RVE}~\cite{moussaddy_assessment_2013}).
In practice, the RVE size is also limited from above by the requirement of separation of scales. When violated, the finite-size bounds can serve only as an input to stochastic finite element calculations~\cite{salmi_various_2012,ostoja-starzewski_random_1998} or higher-order terms have to be introduced in a fully nested numerical homogenization~\cite[and references therein]{kouznetsova_multi-scale_2002,geers_multi-scale_2010,matous_review_2017}.

The RVE size depends on the type of treated physical phenomena, microstructure geometry, and contrast in microstructure constituent properties~\cite{stroeven_numerical_2004}. In the case of high contrast~\cite{dirrenberger_towards_2014} or non-linear behaviour~\cite{stroeven_numerical_2004,gitman_representative_2007,pelissou_determination_2009} the influence of particular geometry gets significantly pronounced, leading in turn to much larger RVE sizes or even to non-existence of an RVE~\cite{gitman_representative_2007}.
Therefore, any recommendation on the RVE size, e.g., those for carbon reinforced polymers made by ~\citet{trias_determination_2006} or for particulate media~\cite[and references therein]{gitman_quantification_2006}, are always highly material-specific and cannot be applied to other materials~\cite{matous_review_2017}. Consequently, similar procedures have to be performed for each investigated material, making the RVE determination still an open topic.

Plenty of works have been devoted to numerical studies of the RVE size; see \Sref{sec:RVEdefinition} for an overview. The prevalent scheme is to (i) generate an ensemble of Statistical Volume Elements (SVEs), i.e., stochastic microstructure realizations smaller than an RVE, and (ii) compute their apparent properties under suitable boundary conditions.
Then, depending on convergence criteria related to distribution of apparent properties within the ensemble, either a new ensemble of larger SVEs is produced or the generated SVEs are declared RVEs for the given threshold. 
The criteria typically involve fluctuations in the apparent properties~\cite{kanit_determination_2003,stroeven_numerical_2004}, their discrepancy under different boundary conditions~\cite{ostoja-starzewski_random_1998}, or a combination of both~\cite{trias_determination_2006,salmi_various_2012,moussaddy_assessment_2013}.
Such an approach is involved because of (i) the need to generate statistically representative microstructure realizations of increasing size and (ii) the computational cost of calculating the apparent properties. For the former reason, most works resort to simple microstructure models, e.g., particulate media~\cite{gusev_representative_1997,segurado_numerical_2002,stroeven_numerical_2004,salmi_apparent_2012,zohdi_method_2001,zohdi_aspects_2001} or Vorono{\"{i}} tessellations~\cite{kanit_determination_2003}.

In this contribution, we address these drawbacks simultaneously by exploring the formalism of Wang tiling, recalled in \Sref{sec:wang_tiling}. Our approach decouples the microstructure generation into the off-line and on-line phases. In the off-line phase, the microstructure is compressed in a set of mutually compatible domains---Wang tiles. During the calculations, microstructure realizations are assembled from the compressed set following a simple on-line stochastic algorithm. As a result, arbitrarily large yet statistically coherent realizations of the compressed microstructure can be generated almost instantly.

Additional advantages of the tiling concept follow from the natural decomposition of the tiling-based microstructure realisations into regular non-overlapping domains. This allows us to employ the Partition theorem by Huet~\cite{huet_application_1990}, revisited in \Sref{sec:partition}, to infer confidence intervals of the homogenized properties by statistical sampling. Moreover, the computational cost of determining apparent properties can be alleviated by standard domain decomposition techniques~\cite{kruis_domain_2006}.

Taking the aforementioned benefits into account, in~\Sref{sec:methodology} we propose a methodology to identify the RVE size for a user-defined accuracy. New microstructure realizations are added on-the-fly to the sequentially generated ensembles of SVEs of increasing size in order to achieve a prescribed confidence in apparent properties, computed with the first-order numerical homogenization recalled in \Sref{sec:homogenization}. The termination criterion, i.e., whether the RVE size has been reached, is based on statistical hypothesis testing related to the provided accuracy, similarly to, e.g.,~\cite{salmi_apparent_2012,saroukhani_statistical_2015}. In \Sref{sec:results}, we apply the proposed methodology to the RVE size determination of three microstructure models: a microstructure with mono-disperse elliptic inclusions, foam, and sandstone.

\paragraph{Scope restrictions}
In what follows, we consider only two-dimensional problems for the sake of clarity, but the extension to three dimensions in the form of Wang Cubes is straightforward~\cite{kari_aperiodic_1995,doskar_aperiodic_2014}. We take existing tiling-based compressions of material microstructures, obtained with methods described in our previous publications, e.g.,~\cite{novak_compressing_2012,doskar_aperiodic_2014}, as fixed inputs capable of sufficiently accurate representation of the microstructure, and perform parametric studies on these geometries.
We also restrict our attention to linearised elasticity and thermal conduction, because an RVE for these linear problems is well defined. In the general case of non-linear models, for instance due to the deterministic size effect in the softening regime, an RVE may not exist in the classical sense~\cite{gitman_representative_2007}, and appropriate modification has to be adopted, e.g., a traction-opening formulation~\cite{verhoosel_computational_2010} or averaging only over active damaging domain~\cite{phu_nguyen_existence_2010}.

\paragraph{Notation}
Throughout the paper, we employ the tensorial notation: scalars are denoted with plain letters, e.g., $\scal{a}$; first- and second-order tensors are typeset with bold italic letters, either $\tens{a}$ or $\tenss{A}$; and fourth-order tensors are written in regular bold letters, $\tensf{A}$.

\section{Wang tiling for random heterogeneous materials}
\label{sec:wang_tiling}

\subsection{Background}
\label{sec:background}
%
%
\begin{figure*}[!ht]
	\centering
	\setlength{\tabcolsep}{0pt}
	\begin{tabular}{>{\centering}m{0.135\textwidth} >{\centering}m{0.335\textwidth} >{\centering}m{0.4300\textwidth} }
		\multicolumn{3}{c}{\includegraphics[width=0.9\textwidth]{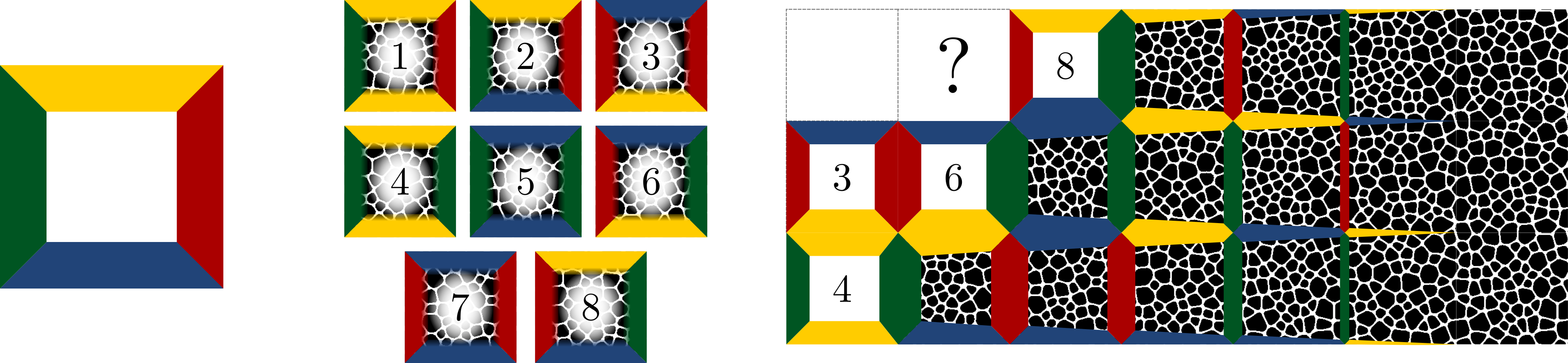}} \\
		(a) & (b) & (c)
	\end{tabular}
	\caption{Illustration of (a) the abstract definition of a Wang tile, (b) a compressed microstructure within a tile set with highlighted tile indices (note that tiles 4, 5, and 7 are self-compatible and, thus, each can stand as a Statistically Equivalent Periodic Unit Cell), and (c) a partially reconstructed microstructure and the underlying tiling, where a step in the stochastic assembly algorithm is depicted. Based on the codes of previously placed tiles 8 and 6, at this step, either tile 2 or 7 will be placed at the position marked with \enquote{?} and the algorithm will continue with the remaining empty position.}
	\label{fig:concept_illustration}
\end{figure*}
The idea of Wang tiling resembles a game of jigsaw puzzle, except that there is only a small set of distinct jigsaw pieces with an infinite number of copies. The jigsaw pieces---\textit{Wang tiles}---have codes attributed to their edges. The goal is to cover a portion of a plane, denoted as a \textit{tiling}, with tile instances from a given \textit{tile set}, such that the adjacent tiles share the same code on the corresponding edge, see~\Fref{fig:concept_illustration}. The codes thus play a role of compatibility constraints during an assembly. Moreover, the tiles can be neither rotated nor reflected during a tiling procedure.

%
The particular set needs to be accompanied by an assembly algorithm capable of producing valid tilings. In our applications, we prefer the stochastic assembly algorithm introduced by \citet{cohen_wang_2003} over deterministic automata, e.g.,~\cite[Chapters 10 and 11]{grunbaum_tilings_1987}, because the former allows for larger variability in design of the tile set. The stochastic assembly algorithm depicted in~\Fref{fig:concept_illustration} works as follows: an empty grid is sequentially filled with tile instances in the row-by-row order, the so called scanline algorithm~\cite{lagae_alternative_2006}. At each step, a tile to-place (denoted with \enquote{?} in~\Fref{fig:concept_illustration}) is randomly chosen from a subset containing only admissible tiles with respect to the codes of the previously places ones. In~\Fref{fig:concept_illustration} the candidate tiles are 2 and 7 as they have red code on the right-hand edge and blue code at the bottom. Randomness of the whole procedure is ensured with presence of at least two tiles for each combination of codes on horizontal and vertical edges~\cite{cohen_wang_2003}. Examples of the tile sets and generated tilings are provided in~\Sref{sec:results}, Figs.~\ref{fig:elliptic_inclusions}, \ref{fig:foam_microstructure}, and \ref{fig:sandstone}.

\subsection{History and relevant applications}
%
The abstract concept of Wang tiling was originally proposed by the mathematician Hao Wang as a semi-decision procedure for proving logical statements of the AEA\footnote{A statement containing two universal and one existential quantifiers.} class~\cite{wang_proving_1961}. His conjecture that an infinite plane can be tiled only periodically was subsequently disproved by \citet{berger_undecidability_1966}, who constructed the first aperiodic set of 20,426 tiles and proposed the corresponding assembly algorithm. Over the years, this number has been reduced~\cite{grunbaum_tilings_1987} down to the currently smallest set of 11 tiles by~\citet{jeandel_aperiodic_2015}, who also conjectured this set to be the smallest possible.

%
Outside discrete mathematics, Wang tiling has found its use in Computer Graphics for an efficient real-time synthesis of a blue noise, which is an essential point distribution for anti-aliasing and dithering, e.g., \cite[and references therein]{kopf_recursive_2006}. In addition, Cohen\etal{}~\cite{cohen_wang_2003} demonstrated that Wang tiles deliver excellent performance in generating naturally looking textures. This development directly motivated our application of Wang tiles in microstructure modelling as discussed next.

\subsection{Tiling in modelling of heterogeneous materials}
%
%
In what follows, a computer model of a material microstructure is understood as a process of generating individual realizations with spatial statistics corresponding to the investigated material~\cite{liu_random_2015}. Pioneered by Povirk~\cite{povirk_incorporation_1995}, most representations are generated with optimization algorithms that minimize discrepancy in statistical characterization of the reference and the generated microstructure. Due to the multi-modal nature of the optimization problem, the simulated annealing, e.g., \cite{yeong_reconstructing_1998,kumar_using_2006,zeman_random_2007}, holds a prominent place among the algorithms; however, other methods such as gradient algorithms~\cite{povirk_incorporation_1995,fullwood_gradient-based_2008}, genetic algorithms~\cite{zeman_random_2007,lee_three-dimensional_2009}, or phase-recovery~\cite{fullwood_microstructure_2008} have been successfully applied.
The microstructural morphology is usually characterised by means of either Minkowski functionals~\cite{scheunemann_design_2015} or set of $n$-point correlation functions~\cite{torquato_random_2002}. In the latter, the two-point probability function~\cite{jiao_modeling_2007,zeman_numerical_2001,rozman_efficient_2001} supplemented either with the two-point cluster~\cite{jiao_superior_2009} or the lineal path~\cite{lu_lineal-path_1992,kumar_using_2006,zeman_random_2007,havelka_compression_2016} functions proved to be sufficient to capture major geometrical features at acceptable computational costs.

Recently, new approaches to microstructure models have emerged, inspired by texture synthesis in Computer Graphics~\cite{wei_fast_2000}, that make use of samples of the reference microstructure. Individual realizations are sequentially generated as a Markovian process with voxels~\cite{liu_random_2015,bostanabad_stochastic_2016} or whole patches~\cite{tahmasebi_cross-correlation_2013} from the reference sample. The suitable voxel/patch values are either chosen according to the statistical proximity of their surrounding in the reference sample to the previously generated portion of the new microstructure realization~\cite{liu_random_2015,tahmasebi_cross-correlation_2013}, or generated by a supervised learning model based on classification trees~\cite{bostanabad_stochastic_2016}.

The common feature of these approaches is that they deliver a statistically similar realization under periodic boundary conditions, which is referred to as a Statistically Optimal Representative Unit Cell~\cite{lee_three-dimensional_2009}, a Statistically Similar Representative Volume Element~\cite{balzani_construction_2014}, or a Statistically Equivalent Periodic Unit Cell (SEPUC)~\cite{zeman_random_2007}. Consequently, each new microstructure realization requires a new, often computationally intensive run of the generating procedure. An alternative is to tile periodically a larger domain with a previously generated cell; however, this introduces long-range periodic artefacts.

From this viewpoint, the stochastic Wang tiling concept presents a compromise between the two aforementioned approaches. The tile set generalizes the notion of SEPUC; instead of being attributed to a single cell, the microstructural information is compressed in a handful of tiles with defined mutual compatibility. While the short-range features of a microstructure are present predominantly in the tile interiors, the long-range characteristics are captured through the particular distribution of the tile edge codes, governing the compatibility requirements. 
Once the microstructure is compressed in the off-line phase, its realizations of any size are generated almost instantly by the assembly algorithm introduced in \mbox{\Sref{sec:background}} (and illustrated with~\mbox{\Fref{fig:concept_illustration}}). In contrast to the above-mentioned periodic extension, the assembled realizations---tilings---are stochastic and exhibit suppressed periodicity artefacts; see~\cite{novak_compressing_2012,doskar_aperiodic_2014}. These features make the microstructure representation based on the Wang tile formalism appealing in applications where multiple (possibly large) microstructure realizations are required.


%

\subsection{Microstructure compression}
\label{sec:compression}
%
%
Microstructure compression amounts to designing a tile set (i.e., its cardinality and distribution of the tile edge codes) and morphologies of tiles within the tile set such that (i) the microstructure remains continuous across the congruent edges and (ii) assembled realizations match the reference microstructure, usually in terms of target spatial statistics. Note that spatial statistics of an individual tile differ from that of the compressed microstructure in general and the proximity of the spatial statistics is required only for assembled tilings.

Because the Wang tile based representation generalizes the SEPUC approach, methods developed for SEPUC can be extended to the generalized periodic boundary conditions appearing in the Wang tiling concept. We have already reported approaches based on optimization algorithms with objective function taking into account the discrepancy in the two-point probability functions~\cite{novak_compressing_2012} or inter-tile traction jumps~\cite{novak_microstructural_2013}. In order to circumvent the computational complexity of the optimization approach, we have also adopted the Computer Graphics approach of~\citet{cohen_wang_2003}, which generates the tile morphology from provided samples of a texture, and we enhanced it with spatial statistics, namely, the two-point probability and cluster functions in~\cite{doskar_aperiodic_2014}. This procedure was used to design the tile set in~\Fref{fig:sandstone}. 

The methodology developed in the following sections holds for arbitrary tile sets, irrespectively of the specific tile design algorithm, providing that the microstructure is accurately captured in the tile set; see~\mbox{\cite[for additional details]{novak_compressing_2012,doskar_aperiodic_2014,doskar_jigsaw_2016}}.


\section{RVE and numerical homogenization}
\label{sec:homogenization}

In this work, we assume the simplest linear constitutive laws at both the micro and the macro scales. Consequently, knowledge of microstructure compositions can be readily propagated to the upper scale by homogenized parameters of an effective constitutive model. The first-order numerical homogenization is summarized in Subsection~\ref{sec:first-order_homogenization}, providing us with boundary-condition biased apparent properties. Next, the notion of RVE is introduced in Subsection~\ref{sec:RVEdefinition}. Finally, Subsections \ref{sec:bounds} and \ref{sec:partition} recall the hierarchy  of bounds and the Partition theorem, respectively, relating the apparent properties of a domain and its subdomains.

\subsection{First-order numerical homogenization}
\label{sec:first-order_homogenization}

\subsubsection{Linear elasticity}
\label{sec:homogenization_LE}

Assume the first-order decomposition of a displacement field $\vardispl$ in the form;
\begin{linenomath}
\begin{equation}
	\vardispl\atx = \tenss{\Strain}\scontr\x + \varfdispl\atx \quad\forall\x\in\domainsize,\varfdispl\!\in\mathcal{U}_{s}^{\mathcal{B}}\,,
	\label{eq:first_order_decomposition}
\end{equation}
\end{linenomath}
where $\tenss{\Strain}$ is the (prescribed) macroscopic strain tensor, $\domainsize\subset\setRn{\text{d}}$ denotes the $\text{d}$-dimensional finite-size domain of a microstructure sample of characteristic size $\svesize$, and $\mathcal{U}_{s}^{\mathcal{B}}$ defines a set of admissible displacement fluctuation fields $\varfdispl$.

For a given $\domainsize$ and $\mathcal{U}_{s}^{\mathcal{B}}$, we define the apparent stiffness tensor $\Dapp$ with the variational equality 
\begin{linenomath}
\begin{equation}
\begin{split}
\tenss{\Strain}\dcontr\Dapp\dcontr\tenss{\Strain}=\\
\inf_{\varfdispl \in \mathcal{U}_{s}^{\mathcal{B}}} \avgsdom{%
	\left(\tenss{\Strain}+\nabla^{\text{s}}\varfdispl\atx\right)
	\dcontr
	\tensf{\D}\atx
	\dcontr
	\left(\tenss{\Strain}+\nabla^{\text{s}}\varfdispl\atx\right)
}{\domainsize}\,,
\label{eq:variational_hom}
\end{split}
\end{equation}
\end{linenomath}
where $\tensf{\D}\atx$ is a local stiffness tensor, $\nabla^{\text{s}}$ stands for the symmetric part of the gradient, and $\avgsdom{\bullet}{\domainsize}$ denotes spatial averaging defined as
\begin{linenomath}
\begin{equation}
	\avgsdom{\bullet\atx}{\domainsize} = \frac{1}{\measure{\domainsize}} \int_{\domainsize} \bullet\atx \de{\tens{x}} \,.
	\label{eq:avareged_quantity_def}
\end{equation} 
\end{linenomath}
The actual strain $\tenss{\strain}$ and stress $\tenss{\stress}$ fields then follow from the minimizer\footnote{For the sake of conciseness, we do not state the explicit dependence $\fdispl=\fdispl\!\left(\tenss{\Strain}\right)$.} $\fdispl$ of~\Eref{eq:variational_hom} through the standard expressions
\begin{linenomath}
\begin{equation}
\tenss{\strain}\atx = \tenss{\Strain} + \nabla^{\text{s}} \fdispl\atx\quad\text{and}\quad 
\tenss{\stress}\atx = \tensf{\D}\atx\dcontr\tens{\strain}\atx\,,
\end{equation}
\end{linenomath}
where we have used the generalized Hooke's law.

Allowing only $\mathcal{U}_{s}^{\mathcal{B}}$ such that $\tenss{\strain}$ and $\tenss{\stress}$ satisfy the energy consistency, also known as Hill's condition~\cite{hill_elastic_1963},
\begin{linenomath}
\begin{equation}
	\avgsdom{\tenss{\stress}\atx\dcontr\tenss{\strain}\atx}{\domainsize} = \avgsdom{\tenss{\stress}\atx}{\domainsize}\!\dcontr\!\avgsdom{\tenss{\strain}\atx}{\domainsize},
	\label{eq:hill_criterion}
\end{equation}
\end{linenomath}
allows us to directly relate $\Dapp$ to the average stress,
\begin{linenomath}
\begin{align}
	\avgsdom{ \tenss{\stress}\atx }{\domainsize} = \avgsdom{ \tensf{\D}\atx \dcontr \tenss{\strain}\atx }{\domainsize} = \Dapp \dcontr \tenss{\Strain} \,, 
	\label{eq:mechanical_def_stiffness}
\end{align}
\end{linenomath}
which will be later used for computing the apparent properties.

%
Posing~\Eref{eq:variational_hom} as a Boundary Value Problem, Hill's criterion is satisfied by adopting $\mathcal{U}_{s}^{\mathcal{B}}$ from the family of Mixed Uniform Boundary Conditions~\cite{hazanov_order_1994},
\begin{linenomath}
\begin{equation}
	\mathcal{U}_{\svesize}^{\textrm{K}} \subseteq \mathcal{U}_{s}^{\mathcal{B}} \subseteq \mathcal{U}_{\svesize}^{\textrm{S}}\,,
	\label{eq:space_ordering}
\end{equation}
\end{linenomath}
where $\mathcal{U}_{\svesize}^{\textrm{K}}$ and $\mathcal{U}_{\svesize}^{\textrm{S}}$ represent the sets of admissible fields $\varfdispl$ compliant with the Kinematic and Static Uniform Boundary Conditions.\footnote{Although the Periodic Boundary Conditions usually provide reasonable estimates of the homogenized properties even in the case of a non-periodic microstructure~\cite{michel_effective_1999,kanit_determination_2003,sab_periodization_2005}, our RVE criterion is based on the discrepancy between the bounds on the apparent properties. Therefore, we omit discussion on Periodic Boundary Conditions onwards.} The particular forms are specified as follows:

\paragraph[Kinematic Uniform Boundary Conditions (KUBC)]{Kinematic Uniform Boundary Conditions (KUBC)\rempunct} impose a prescribed displacement at the domain boundary $\boundarysize$ in the form
\begin{linenomath},
\begin{equation}
	\vardispl\atx = \tenss{\Strain} \scontr \tens{x} \quad \forall\tens{x} \in \boundarysize \,,
	\label{eq:KUBC_condition_displ}
\end{equation}
\end{linenomath}
resulting in vanishing fluctuation displacements at $\boundarysize$. This corresponds to setting
\begin{linenomath}
\begin{equation}
	\mathcal{U}_{s}^{\textrm{K}} = \left\{ \varfdispl:\domainsize\to\setRn{\text{d}}; \varfdispl|_{\boundarysize} = \tens{0} \right\}\,.
	\label{eq:KUBC_condition_fluctu}
\end{equation}
\end{linenomath}

\paragraph[Static Uniform Boundary Conditions (SUBC)]{Static Uniform Boundary Conditions (SUBC)\rempunct} are traditionally defined with affine traction vectors at $\boundarysize$, leading to a stress-controlled problem. However, \citet{miehe_computational_2003} proved that SUBC correspond to the so-called minimal Kinematic Boundary Conditions, used in e.g.~\cite{mesarovic_minimal_2005,gluge_generalized_2013,doskar_jigsaw_2016}, that require
\begin{linenomath}
\begin{equation}
	\tenss{\Strain}=\avgsdom{\nabla^{\text{s}}\vardispl\atx}{\domainsize}\!.
	\label{eq:SUBC_condition_displ}
\end{equation}
\end{linenomath}
Similarly to \Eref{eq:KUBC_condition_fluctu}, this provides the specific form of $\mathcal{U}_{s}^{\mathcal{B}}$ as
\begin{linenomath}
\begin{equation}
\begin{split}
	\mathcal{U}_{s}^{\textrm{S}} = \{\varfdispl\!:\domainsize\!\to\setRn{\text{d}}; \int_{\partial\domainsize}\!\!\tens{n}\otimes\varfdispl\de{\Gamma} = \tenss{0},\,
	\avgsdom{\varfdispl\atx}{\domainsize}\!\!=\tens{0}\}\,.
	\label{eq:SUBC_condition_fluctu}
\end{split}
\end{equation}
\end{linenomath}
Note that the boundary integral contains also non-symmetric part of the gradient, which along with the last condition in~\Eref{eq:SUBC_condition_fluctu} prevents rigid body modes.

\subsubsection{Thermal conduction}
\label{sec:homogenization_thermal}
For thermal conduction, we can proceed analogously to linear elasticity with only minor modifications: Generalized Hooke's law is replaced with Fourier's law, $\tens{\flux}\atx = - \tenss{\K}\atx \scontr\tens{\tempgrad}\atx$, which governs the relation between a heat flux $\tens{\flux}$ and a temperature gradient $\tens{\tempgrad} = \nabla \scal{\tilde{\temp}}$ via a thermal conductivity tensor $\tenss{\K}$. The first order decomposition of an admissible temperature field $\tilde{\scal{\temp}}\atx$ reads
\begin{linenomath}
\begin{equation}
	\tilde{\temp}\atx = \tens{\Tempgrad} \scontr \x + \tilde{\scal{\temp}}^{*}\atx  \quad \forall \tens{x} \in \domainsize, \tilde{\scal{\temp}}^{*}\in\mathcal{T}_{s}^{\mathcal{B}}\,,
	\label{eq:temperature_decomposition}
\end{equation}
\end{linenomath}
with $\tens{\Tempgrad}$ denoting the prescribed macroscopic temperature gradient and $\mathcal{T}_{s}^{\mathcal{B}}$ being, again, the set of admissible temperature fluctuation fields compliant with Hill's condition
\begin{linenomath}
\begin{equation}
	\avgsdom{ \tens{\flux}\atx \scontr \tens{\tempgrad}\atx }{\domainsize} = \avgsdom{\tens{\flux}\atx}{\domainsize} \scontr \avgsdom{\tens{\tempgrad}\atx}{\domainsize}\!.
	\label{eq:hill_lemma_TR}
\end{equation}
\end{linenomath}

Consequently, the variational definition of the apparent conductivity tensor $\Kapp$,
\begin{linenomath}
\begin{equation}
\begin{split}
	\tens{\Tempgrad}\scontr\Kapp\scontr\tens{\Tempgrad}=\\
	\inf_{{\tilde{\temp}}^{*}\in\mathcal{T}_{s}^{\mathcal{B}}} \langle \tenss{\Tempgrad}+\nabla\tilde{\temp}^{*}\atx) \scontr \tenss{\K}\atx \scontr ( \tenss{\Tempgrad}+\nabla\tilde{\temp}^{*}\atx) \rangle_{\domainsize}\,,
	\label{eq:energetical_def_resistance}
\end{split}
\end{equation}
\end{linenomath}
is equivalent to the volume averaging of the heat flux obtained from the minimizer $\temp\atx$ of~\Eref{eq:energetical_def_resistance}
\begin{linenomath}
\begin{equation}
	\avgsdom{\tens{\flux}\atx}{\domainsize} = \avgsdom{-\tenss{\K}\atx\scontr\nabla\temp\atx}{\domainsize}=-\Kapp\scontr\tens{\Tempgrad}\,.
	\label{eq:mechanical_def_conductivity}
\end{equation}
\end{linenomath}

As in the previous section, \Eref{eq:hill_lemma_TR} can be ensured with a proper choice of $\mathcal{T}_{s}^{\mathcal{B}}$ that falls within the following two limit cases:

\paragraph[Uniform Temperature Gradient Boundary Conditions]{Uniform Temperature Gradient Boundary Conditions\rempunct} prescribing values at the boundary in the form
\begin{linenomath}
\begin{equation}
	\scal{\temp}\atx = \tens{\Tempgrad} \scontr \tens{x} \quad \forall \tens{x} \in \boundarysize \,,
	\label{eq:UGBC_condition_temp}
\end{equation}
\end{linenomath}
which translates to
\begin{linenomath}
\begin{equation}
	\mathcal{T}_{\svesize}^{\text{G}} = \{\tilde{\temp}^{*}:\domainsize\to\setR;\,\tilde{\temp}^{*}|_{\boundarysize}=0\}\,,
\end{equation}
\end{linenomath}	 
and
\paragraph[Uniform Heat Flux Density Boundary Conditions]{Uniform Heat Flux Density Boundary Conditions\rempunct}defined analogously to~\Eref{eq:SUBC_condition_displ} and represented by
%
\begin{linenomath}
\begin{equation}
	\mathcal{T}_{\svesize}^{\textrm{F}} = \{\tilde{\temp}^{*}:\domainsize\to\setR; \int_{\boundarysize}\!\!\scal{\tilde{\temp}^{*}}\tens{n}\de{\Gamma} = \tens{0},\,
	\avgsdom{\scal{\tilde{\temp}^{*}}\atx}{\domainsize}\!\!=0\}\,.
\end{equation}
\end{linenomath}


\subsection{Notion of RVE}
\label{sec:RVEdefinition}

%
The apparent properties introduced above are in general boundary condition biased and individual realizations of the microstructure yield different tensors, which contradicts the original requirement of Hill~\cite{hill_elastic_1963}. However, Sab~\cite{sab_homogenization_1992} proved that boundary-biased apparent properties converge to the homogenized ones with increasing size \svesize, thus
\begin{linenomath}
\begin{equation}
	 \Dapp\xrightarrow{s\rightarrow\infty}\Dhom \quad\mathrm{and}\quad \Kapp\xrightarrow{s\rightarrow\infty}\Khom \,,
	\label{eq:pointwise_convergence}
\end{equation}
\end{linenomath}
see also~\citet{bourgeat_approximations_2004}.
%


%
As discussed in Introduction, the theoretical RVE is conventionally replaced with a finite size numerical counterpart~\cite{moussaddy_assessment_2013}. The common approach to the determination of the numerical RVE size rests on generating ensembles containing sequentially larger SVEs until statistics of the obtained data comply with a user-defined threshold. The investigated data ranges from classical overall stiffness parameters~\cite{terada_simulation_2000,shan_representative_2002,trias_determination_2006,salmi_various_2012} and elastic strain energy density~\cite{saroukhani_statistical_2015}, to mean values or concentrations in microstructural stress and strain fields~\cite{shan_representative_2002,stroeven_numerical_2004,gitman_quantification_2006,gitman_representative_2007,trias_determination_2006}, to deviations from assumed macroscopic isotropy~\cite{moussaddy_assessment_2013,salmi_apparent_2012}.
Statistical control usually includes output variance for different realizations of the same size and convergence of the mean value from one SVE size to another. Relying on a single criterion, especially when combined only with one type of boundary conditions (e.g., Periodic Boundary Conditions), can lead to pre-mature convergence~\cite{moussaddy_assessment_2013}; therefore, more recent works combine both characteristics~\cite{trias_determination_2006,salmi_various_2012,moussaddy_assessment_2013,saroukhani_statistical_2015}.

%
In order to alleviate the computational cost related to the above mentioned approach, Kanit\etal{} in their seminal work~\cite{kanit_determination_2003} adopted the notion of the integral range allowing them to establish a power-law relation among an SVE size, cardinality of an ensemble, and the variation of apparent properties. Parameters of the relation are calibrated with only a handful of computations and the RVE size is then derived with respect to a user-defined statistical variation threshold. Moreover, the expression also allows to substitute a single RVE with a set of smaller SVEs. Pelissou\etal{}~\cite{pelissou_determination_2009} enhanced the original approach by introducing uncertainty to both the mean value and the variance, using the bootstrapping method, and applied it to non-linear problems. Recently, Dirrenberger\etal{}~\cite{dirrenberger_towards_2014} extended Kanit et al.'s approach to an artificial microstructure with infinite integral range demonstrating that the approach is applicable even for highly complex materials. The method of \citet{kanit_determination_2003} has also been successfully applied to the real world tasks arising, for instance, in the food industry~\cite{kanit_apparent_2006,kanit_virtual_2011}.
However, the variance based criterion of~Kanit\etal{} relies on an implicit assumption that the mean value is not significantly biased by the prescribed boundary conditions. This assumption is usually valid for large SVEs under Periodic Boundary Conditions~\cite{michel_effective_1999,sab_periodization_2005}, but the assumption becomes questionable for complex microstructures with high contrast in phase properties~\cite{dirrenberger_towards_2014}. 

A different approach to an RVE definition was proposed by Drugan and Willis~\cite{drugan_micromechanics-based_1996}. Estimating the effect of strain average fluctuations in a non-local constitutive equation allowed them to derive the RVE size of two particle diameters for a microstructure composed of non-overlapping spheres. Their analytical findings were later corroborated in numerical studies of Gusev~\cite{gusev_representative_1997} and Sequrado and Llorca~\cite{segurado_numerical_2002}. Another alternative definition of an RVE has been recently proposed by Hoang\etal{}~\cite{hoang_determining_2016}, who combined incremental analytical and numerical homogenizations. Their RVE criterion rests on the convergence of parameters in the analytical homogenization identified to follow results of the numerical homogenization.

Some authors, e.g.,~\cite{shan_representative_2002,trias_determination_2006,niezgoda_optimized_2010}, also incorporated additional statistics into their definition of the RVE. Following Kanit et al.'s idea of replacing a single RVE with a set of smaller ones, \citet{niezgoda_optimized_2010} introduced the notion of RVE Set composed of optimally chosen SVEs from an ensemble whose convex combination best matches the ensemble average of given microstructural statistics, namely, the two-point correlation function. The optimal convex combination is then used for computing all macroscopic properties.


\subsection{Bounds on the apparent properties}
\label{sec:bounds}

%
For the sake of conciseness, we recall the hierarchy of bounds for linear elasticity only; however, the exposition can be straightforwardly applied also to the problem of thermal conduction. 

In linear elasticity, KUBC and SUBC hold a prominent place as they provide bounds on the apparent property~\cite{hill_elastic_1963,huet_application_1990}
\begin{linenomath}
\begin{equation}
	\tensf{\D}_{\svesize}^{\textrm{S}} \preccurlyeq \Dapp \preccurlyeq  \tensf{\D}_{\svesize}^{\textrm{K}} \,,
	\label{eq:apparent_bounds}
\end{equation}
\end{linenomath}
with the ordering relation $\preccurlyeq$ defined for fourth-order tensors $\tensf{A}$ and $\tensf{B}$ in the sense
\begin{linenomath}
\begin{equation}
	\tensf{A} \preccurlyeq \tensf{B} \iff \tenss{a} \dcontr \left( \tensf{B} - \tensf{A} \right) \dcontr \tenss{a} \geq 0 \quad \forall \tenss{a} \in \setR^{\textrm{d}\times\textrm{d}} \,.
\end{equation}
\end{linenomath}
This classical ordering directly follows from the principle of minimum potential energy,~\Eref{eq:variational_hom}, and the definition of the kinematically admissible spaces, \Eref{eq:space_ordering}.
%

%
Beside the realization-to-realization convergence of the apparent properties to the homogenized ones, Sab~\cite{sab_homogenization_1992} also proved that the homogenized properties can be bounded by ensemble averages. In particular, it holds
%
\begin{linenomath}
\begin{equation}
	\left(\sup_{\svesize} \mathbb{E}\left( \tensf{\C}_{\scal{\svesize}}^{\textrm{S}} \right)\right)^{-1}\!\!= \tensf{D}^{\homg}\! = \inf_{\svesize} \mathbb{E}( \tensf{\D}_{\scal{\svesize}}^{\textrm{K}} )\,,
	\label{eq:infinite_hierarchy}
\end{equation}
\end{linenomath}
where $\mathbb{E}(\bullet)$ denotes the expected value of an ensemble average over all microstructure realizations of the same size and $\tensf{\C}_{\scal{\svesize}}^{\textrm{S}} = \left(\tensf{\D}_{\scal{\svesize}}^{\textrm{S}}\right)^{-1}$.

\subsection{Partition Theorem}
\label{sec:partition}
%
In the case of a finite-size domain, a hierarchy of bounds similar to Eqs.~(\ref{eq:apparent_bounds}) and (\ref{eq:infinite_hierarchy}) can be established for the apparent properties of the domain and its subdomains. This was first recognized by Huet~\cite{huet_application_1990} under the name Partition theorem with implications for physical testing of materials whose representative volumes are unattainable for practical experiments. Later, the same hierarchy appeared in~\cite{sab_homogenization_1992} as the sub-additivity property of apparent tensors, and in~\mbox{\cite[and subsequent works]{zohdi_domain_1999}} as a consequence of error bounds in the substructuring method~\cite{zohdi_method_2001}.

\begin{figure}
	\centering
	\includegraphics[width=\columnwidth]{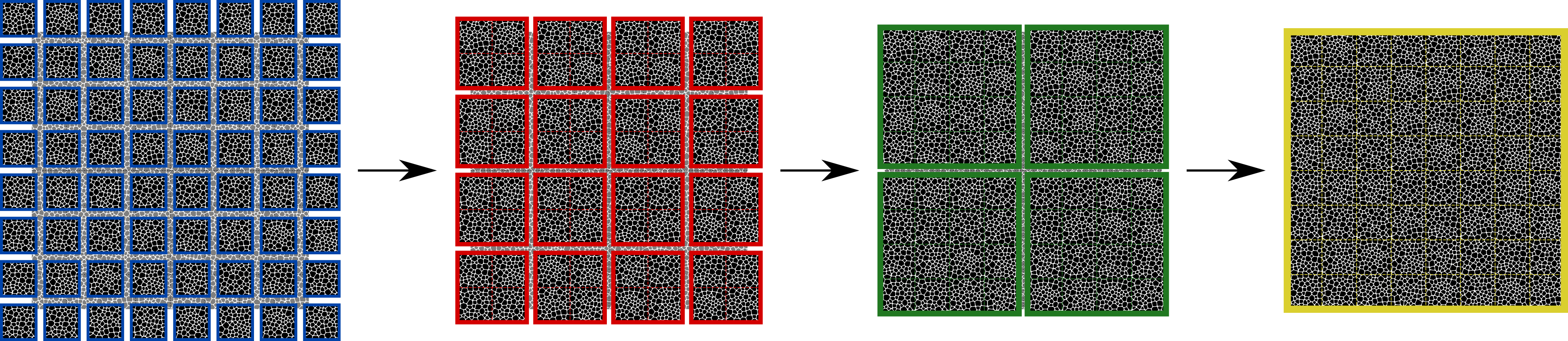}
	\caption{Illustration of the Partition theorem: The apparent properties {\color[rgb]{0.854901,0.811764,0.184314}${\Dtens_{\svesize_{t+4}}^{\textrm{K}}}$} and {\color[rgb]{0.854901,0.811764,0.184314}${\Ctens_{\svesize_{t+4}}^{\textrm{K}}}$} of the rightmost, yellow-bordered tiling are consecutively bounded by the hierarchies {\color[rgb]{0,0.26667,0.6667}$\overline{\Dtens_{\svesize_{t}}^{\textrm{K}}}$} $\succcurlyeq$ {\color[rgb]{0.83137,0,0}$\overline{\Dtens_{\svesize_{t+1}}^{\textrm{K}}}$} $\succcurlyeq$ {\color[rgb]{0.129411,0.470588,0.129411}$\overline{\Dtens_{\svesize_{t+2}}^{\textrm{K}}}$} $\succcurlyeq$ {\color[rgb]{0.854901,0.811764,0.184314}${\Dtens_{\svesize_{t+4}}^{\textrm{K}}}$} and {\color[rgb]{0,0.26667,0.6667}$\overline{\Ctens_{\svesize_{t}}^{\textrm{S}}}$} $\succcurlyeq$ {\color[rgb]{0.83137,0,0}$\overline{\Ctens_{\svesize_{t+1}}^{\textrm{S}}}$} $\succcurlyeq$ {\color[rgb]{0.129411,0.470588,0.129411}$\overline{\Ctens_{\svesize_{t+2}}^{\textrm{S}}}$} $\succcurlyeq$ {\color[rgb]{0.854901,0.811764,0.184314}${\Ctens_{\svesize_{t+4}}^{\textrm{S}}}$} of ensemble averages of consecutively smaller subdomains of side length $\svesize_{k}$.}
	\label{fig:partition_theorem}
\end{figure}

Assume partitioning of the domain $\domain_{\svesize}$ into $n$ equi-sized non-overlapping subdomains $\domain^{(i)}_{r}, i = 1\dots n,$ of the characteristic size $r<\svesize$.
By solving the variational problem~(\ref{eq:variational_hom}) independently for each subdomain under KUBC, we obtain the collection of solutions
$\left\{\tens{u}^{(i)}\atx, \tens{x} \in \domain^{(i)}_{r}\right\}$. Clearly, a displacement field $\bar{\tens{u}}$ defined for the whole domain $\domainsize$ as
\begin{linenomath}
\begin{equation}
	\bar{\tens{u}}\atx = \left\{ \tens{u}\atx: \domainsize \to \setR^{\textrm{d}}; \left.\tens{u}\right|_{\domain^{(i)}_{r}} = \tens{u}^{(i)} \right\}
\end{equation}
\end{linenomath}
is an admissible field satisfying~\Eref{eq:KUBC_condition_displ}. Hence, plugging $\bar{\tens{u}}$ in~\Eref{eq:variational_hom} leads to
\begin{linenomath}\begin{equation}
	\begin{split}
		\tenss{\Strain} \dcontr \tensf{\D}_{\svesize}^{\textrm{K}} \dcontr \tenss{\Strain} 
		&\leq \avgsdom{ \nabla^{\textrm{s}}\bar{\tens{u}}\atx \dcontr \tensf{D}\atx \dcontr \nabla^{\textrm{s}}\bar{\tens{u}}\atx }{\domainsize} \\ 
		&= \sum_{i}^{n} \frac{\measure{\domain^{(i)}_{r}}}{\measure{\domainsize}} \avgsdom{ \nabla^{\textrm{s}}\tens{u}^{(i)}\atx \dcontr \tensf{D}\atx \dcontr \nabla^{\textrm{s}}\tens{u}^{(i)}\atx }{\domain^{(i)}_{r}} \\ 
		&= \frac{1}{n} \sum_{i}^{n} \tenss{\Strain} \dcontr \tensf{\D}_{r}^{\textrm{K},(i)} \dcontr \tenss{\Strain} \,, \forall \tenss{\Strain} \in \setR_{\textrm{sym}}^{\textrm{d}\times\textrm{d}}\,,
	\end{split}
	\label{eq:partition_theorem_semiproof}
\end{equation}
\end{linenomath}
which provides us with the relation 
\begin{linenomath}
\begin{equation}
	\tensf{\D}_{\svesize}^{\textrm{K}} \preccurlyeq \frac{1}{n} \sum_{i}^{n} \tensf{\D}_{r}^{\textrm{K},(i)} = \overline{\Dtens_{r}^{\textrm{K}}} \,.
\end{equation}
\end{linenomath}
From the Principle of minimum complementary energy, an analogous result for SUBC follows in the form
%
\begin{linenomath}
\begin{equation}
	\tensf{\C}_{\svesize}^{\textrm{S}} \preccurlyeq \frac{1}{n} \sum_{i}^{n} \tensf{\C}_{r}^{\textrm{S},(i)} = \overline{\Ctens_{r}^{\textrm{S}}} \,.
\end{equation}
\end{linenomath}
By recursive partitioning of subdomains and making use of~\Eref{eq:apparent_bounds} and the convergence property~(\ref{eq:pointwise_convergence}), the final hierarchy of bounds can be established
\begin{linenomath}
\begin{equation}
	\begin{split}
		\left( \overline{\Ctens_{0}^{\textrm{S}}} \right)^{-1} &\preccurlyeq \dots \preccurlyeq \left( \overline{\Ctens_{\svesize_{k}}^{\textrm{S}}} \right)^{-1} \preccurlyeq \left( \overline{\Ctens_{\svesize_{k+1}}^{\textrm{S}}} \right)^{-1} \preccurlyeq \dots \preccurlyeq \\
		\Dtens^{\homg} &\preccurlyeq \dots \preccurlyeq \overline{\Dtens_{\svesize_{k+1}}^{\textrm{K}}} \preccurlyeq \overline{\Dtens_{\svesize_{k}}^{\textrm{K}}} \preccurlyeq \dots \preccurlyeq \overline{\Dtens_{\svesize_{0}}^{\textrm{K}}} \,,
	\end{split}
	\label{eq:bounds_hierarchy}
\end{equation}
\end{linenomath}
where $\svesize_k = \frac{1}{q} \svesize_{k+1}$ with $q \in \setN$; see~\Fref{fig:partition_theorem} for an illustration with $q=2$. Note that the outermost bounds $\overline{\Dtens_{0}^{\textrm{K}}}$ and $\left( \overline{\Ctens_{0}^{\textrm{S}}} \right)^{-1}$, obtained as the limit states for $s\to0$, are the classical Voigt and Reuss bounds which are derived under the assumption of homogeneous strain and stress field within the sample.


\section{RVE size determination}
\label{sec:methodology}

%
Equipped with a procedure for computing apparent properties and with a realistic microstructure generation that also provides partitioning in the spirit of Huet~\cite{huet_application_1990}, we formulate a two-level method for identification of the RVE size for a user-defined tolerance and confidence level.

%
Similar ideas have been presented in several works; here, we attempt to encompass the best of these in a comprehensive yet straightforward framework. We build on two assumptions:
\begin{itemize}
	\item statistical homogeneity and ergodicity of the microstructure itself to ensure existence of an RVE~\cite{sab_homogenization_1992};
	\item sufficiently accurate compression of the microstructural information in the form of a Wang tile set. We assume that all \replace{important}{essential} features of the investigated microstructure are present in reconstructed samples\footnote{Albeit reduced compared to the periodic extension of a SEPUC, reconstructed realizations exhibit secondary peaks in spatial correlation functions~\cite{novak_compressing_2012,novak_microstructural_2013,doskar_aperiodic_2014}. These peaks are further reduced in solutions to homogenization-related Boundary Value Problems and averaged; therefore, the influence of the secondary peaks on the apparent properties and the RVE size is marginal.}\replace{}{ and hence the RVE size identified for the tile-based compression corresponds to the RVE size of the microstructure}.
\end{itemize}
Note that the latter assumption is inherently present in any microstructure compression technique, including the SEPUC approach~\cite{zeman_random_2007,niezgoda_optimized_2010}.

Our approach resembles the work of Saroukhani\etal{}~\cite{saroukhani_statistical_2015}, especially in deriving bounds on the homogenized properties with methods of statistical sampling. General bounds on statistical moments were also presented in~\cite{zohdi_statistical_2005,zohdi_introduction_2008}. Contrary to the aforementioned works, which does not provide any quantitative definition of the RVE, we introduce an RVE criterion that is based on hypothesis testing similar to, e.g.,~\cite{gitman_quantification_2006,trias_determination_2006}. Unlike the latter works, the number of microstructure realizations is not defined a priori in our approach, as we control their number on-the-fly in order to meet a prescribed confidence in bounds on the apparent properties.

%
The key idea is to relate the theoretical RVE to an infinite tiling. Any finite-size tiling can thus be considered as a subdomain of the RVE. The bounds in~\Eref{eq:bounds_hierarchy} then implicitly contain infinite sums. Therefore, at the first level of our methodology, we identify the minimal number of microstructure realizations that delivers the bounds with a user-defined uncertainty.
At the second level, we assess the discrepancy between the bounds and, based on statistical hypothesis testing, we decide whether the actual size of microstructure is the RVE size for the defined tolerance.

\subsection{Level I: Bounds for apparent properties}

%
To keep the exposition concise, we adopt a certain abuse of notation in the sequel: $\Ltens$ stands for apparent tensors rendered by prescribing the zero fluctuation unknowns point-wise at the boundary (i.e., $\Ltens = \tensf{D}^{\textrm{K}}$ in the case of linear elasticity and $\Ltens = \tenss{K}^{\textrm{G}}$ for the thermal conduction), whilst $\Mtens$ denotes the complementary quantity obtained by enforcing the zero fluctuations in the weak, boundary-integral sense ($\Mtens = (\tensf{D}^{\textrm{S}})^{-1}$ or $\Mtens = (\tenss{K}^{\textrm{F}})^{-1}$).
Analogously to the linear elasticity problem, the superscripts $\bullet^\textrm{G}$ and $\bullet^\textrm{F}$ denote the apparent conductivity tensors obtained under Uniform Gradient Boundary Conditions and Uniform Flux Boundary Conditions, respectively, recall~\Sref{sec:homogenization_thermal}.

For each realization $\domain^{(i)}_{s}$ of a $s\times s$ tiling\footnote{Size of a realization is always an integer multiple of the tile size.} we define two scalar values
\begin{linenomath}
\begin{equation}
	\LscalOne = \norm{\LtensOne} \quad \textrm{and} \quad \MscalOne = \norm{\MtensOne} \,,
	\label{eq:scalar_quantification}
\end{equation}
\end{linenomath}
that are used to quantify variability of a $s$-size SVE ensemble.
The particular type of the norm in~\Eref{eq:scalar_quantification} is a modelling choice to be made by a user. 
We assume the obtained data to be in the form
\begin{linenomath}
\begin{equation}
	\begin{split}
		\LscalOne &= \LscalHom + \LscalBias + \LscalError = \LscalMean + \LscalError \,,\\
		\MscalOne &= \MscalHom + \MscalBias + \MscalError = \MscalMean + \MscalError \,,\\
	\end{split}
	\label{eq:statistical_form}
\end{equation}
\end{linenomath}
where $\LscalHom$ and $\MscalHom$ correspond to the norms of the sought homogenized tensors $\LtensHom$ and $\MtensHom=(\LtensHom)^{-1}$, $\LscalBias$ and $\MscalBias$ denote the systematic bias caused by specific boundary conditions and the finite size of the domain, and $\LscalError$ and $\MscalError$ are random, normally distributed errors from $N(0,(\LscalErrorStd)^2)$ and $N(0,(\MscalErrorStd)^2)$ related to the stochastic nature of the microstructure. Recall that the systematic errors and the variances $(\std_{\svesize}^\bullet)^2$ of the random errors vanish as
$\svesize\to\infty$, \Eref{eq:pointwise_convergence}.

%
In~\Eref{eq:statistical_form} and in the sequel, the bar-denoted quantities $\scalMean{\bullet}$ represent the theoretical mean value, obtained by averaging the quantity over all possible realizations. Due to the assumed infinite size of the theoretical RVE, these values are unattainable and must be replaced with confidence intervals. To this purpose, for both quantities $\Lscal$ and $\Mscal$ of $\real$ realizations of $\svesize\times\svesize$ tiling, we compute the sample mean values
\begin{linenomath}
\begin{equation}
	\sampleAver{\bullet} = \frac{1}{\real} \sum_{i=1}^{\real} \scalOne{\bullet}
	\label{eq:mean_definition}
\end{equation}
\end{linenomath}
and the unbiased sample standard deviations
\begin{linenomath}
\begin{equation}
	\sampleStd{\bullet} = \sqrt{ \frac{1}{\real - 1} \sum_{i=1}^{\real} \left( \scalOne{\bullet} - \sampleAver{\bullet} \right)^{2} } \,,
	\label{eq:std_definition}
\end{equation}
\end{linenomath}
with $\bullet$ denoting either $\Lscal$ or $\Mscal$. From the Central Limit Theorem, the deterministic value $\scalMean{\bullet}$ falls with $(1-\confLevelOne) \times 100\%$ probability within the confidence interval
\begin{linenomath}
\begin{equation}
	\scalMean{\bullet} \in \left[ \sampleAver{\bullet} - \alpha^{\bullet}_{\svesize}; \sampleAver{\bullet} + \alpha^{\bullet}_{\svesize} \right]\,,
	\label{eq:conf_int}
\end{equation}
\end{linenomath}
with the width of the interval given by
\begin{linenomath}
\begin{equation}
	\alpha^{\bullet}_{\svesize} = \tdist{1-\confLevelOne/2}{\real-1} \frac{\sampleStd{\bullet}}{\sqrt{\real}}\,,
	\label{eq:conf_int_half-width}
\end{equation}
\end{linenomath}
where $\tdist{P}{n}$ denotes the inverse cumulative distribution function of Student's t-distribution and $\confLevelOne$ is a significance level provided by the user for Level I.

%

%
The ratio
\begin{linenomath}
\begin{equation}
	\errorSize^{\bullet}_{\svesize} = \frac{\alpha^{\bullet}_{\svesize}}{\sampleAver{\bullet}}\,
\end{equation}
\end{linenomath}
provides a natural uncertainty measure in the bounds. Note that, due to the presence of $\tdist{1-\confLevelOne/2}{\real-1}$ in~\Eref{eq:conf_int_half-width}, the ratio does not correspond to the (biased) estimation of the coefficient of variation (CoV) used in, e.g.,~\cite{salmi_various_2012,trias_determination_2006}. This complies with our intention to assess the uncertainty of ensemble mean value determination rather than the variation inherent to the limited realization size and the imposed BC.

Microstructure realizations are being added on-the-fly to the ensemble of size $\svesize$ samples until the uncertainty in both upper and lower bounds drops below a given threshold, i.e., $ \errorSize^{\bullet}_{\svesize} < \errorSizeLim$, which translates to asserting that the actual mean value $\scalMean{\bullet}$ falls outside the interval $\interval{(1-\errorSizeLim)\sampleAver{\bullet}}{(1+\errorSizeLim)\sampleAver{\bullet}}$ with less than probability $\confLevelOne$. Once this condition is satisfied, we assume that the ensemble contains sufficient number of realization to provide the desired accuracy for the RVE size criterion, controlled next.

\subsection{Level II: RVE size criterion}

Proximity of each realization to the RVE size is assessed using a discrepancy between $\LtensOne$ and $(\MtensOne)^{-1}$. Recall that $\LtensOne$ and $\MtensOne$ are reciprocal in the RVE case, see~\Eref{eq:infinite_hierarchy}. For each realization, we define the proximity error as
\begin{linenomath}
\begin{equation}
	\errorProximityOne =  \norm{ \LtensOne \scontr \MtensOne - \tenss{I}}\,,
	\label{eq:proximity_error}
\end{equation}
\end{linenomath}
where $\scontr$ denotes the corresponding tensorial contraction\footnote{A single contraction for conductivity and a double contraction for the elasticity problem.} and $\tenss{I}$ is the corresponding unit tensor (with the pertinent symmetries). Again, the particular type of the norm is a modelling choice; for instance, \citet{sab_homogenization_1992} used the infinity norm.

Finally, the RVE size criterion is based on testing the hypothesis 
\begin{linenomath}
\begin{equation*}
\textrm{H}_{0}: \  \scalMean{\errorProximity} \geq \errorProximityLim \quad \text{against} \quad
\textrm{H}_{1}: \  \scalMean{\errorProximity} < \errorProximityLim \,,
\end{equation*}
\end{linenomath}
where $\errorProximityLim$ is a given threshold discrepancy defining the computational RVE. This results in the one-tailed hypothesis test
\begin{linenomath}
\begin{equation}
	 \sampleAver{\errorProximity} + \tdist{1-\confLevelTwo}{\real-1} \frac{\sampleStd{\errorProximity}}{\sqrt{\real}} = \errorProximityTest \leq \errorProximityLim  \,.
	 \label{eq:RVE_criterion}
\end{equation}
\end{linenomath}
If the condition~(\ref{eq:RVE_criterion}) is satisfied for user-defined $\confLevelTwo$ and $\errorProximityLim$, the current tiling size $s$ is declared to be the computational RVE size; otherwise, we proceed with an ensemble of larger tilings.

The proposed methodology is summarized in Algorithm~\ref{alg:methodology}. For practical purposes, size and number of realizations are limited with $\svesizemax$ and $\realmax$, respectively. Moreover, the first $\realmin$ realizations of each size are generated and their apparent properties $\LtensOne$ and $\MtensOne$ computed without comparing $\errorSize^{\bullet}_{\svesize}$ to the threshold value, in order to acquire reliable data statistics for the RVE size criterion. Characteristics of $1\times1$ tilings are computed independently beforehand, because they correspond to the properties of individual tiles weighted by the probability of occurrence of each tile in an infinite tiling, which follows directly form the definition of the tile set, cf. the idea of RVE Sets~\cite{niezgoda_optimized_2010}.

\begin{algorithm}[!ht]
	\begin{algorithmic}
		\Require $\errorSizeLim$, $\confLevelOne$, $\errorProximityLim$, and $\confLevelTwo$
		\State $\svesize \gets 1$
		\Comment{Skip $1\times1$ tilings}
		\Repeat
		\State $\svesize \gets \svesize + 1$
		\State $i \gets 0$
		\Comment{Level I}
		\For{ $i < \realmin$ }
		\State $i \gets i + 1$
		\State generate a tiling $\svesize \times \svesize$
		\State synthesize a microstructure from the tiling
		\State compute $\LtensOne$ and $\MtensOne$
		\State compute $\LscalOne$ and $\MscalOne$
		\EndFor
		\State compute ensemble statistics $\sampleAver{\Lscal}$, $\sampleAver{\Mscal}$, $\sampleStd{\Lscal}$, and $\sampleStd{\Mscal}$
		\State compute $\errorSize^{\Lscal}_{\svesize}$ and $\errorSize^{\Mscal}_{\svesize}$
		
		\While{ $ \left( \errorSize^{\Lscal}_{\svesize} > \errorSizeLim \right) \wedge \left( \errorSize^{\Mscal}_{\svesize} > \errorSizeLim \right) \wedge \left( i < \realmax \right)$ }
		\State $i \gets i+1$
		\State generate a tiling $\svesize \times \svesize$
		\State synthesize a microstructure from the tiling
		\State compute $\LtensOne$ and $\MtensOne$
		\State compute $\LscalOne$ and $\MscalOne$
		\State update ensemble statistics
		\State compute $\errorSize^{\Lscal}_{\svesize}$ and $\errorSize^{\Mscal}_{\svesize}$
		\EndWhile
		\State $\real \gets i$
		\State compute $\sampleAver{\errorProximity}$ and $\sampleStd{\errorProximity}$
		\Comment{Level II}		
		\State compute $\errorProximityTest$
		\Until{ $\left(\errorProximityTest \leq \errorProximityLim\right) \, \vee \, \left( \svesize \geq \svesizemax \right)$ }
	\end{algorithmic}
	\caption{Identification of the numerical RVE size}
	\label{alg:methodology}
\end{algorithm}

\subsection{Alleviating computational cost}
\label{sec:domain_decomposition}

High computational cost is a common sore of procedures aimed at identifying the RVE size. We exploit the repeating occurrence of individual tiles in the microstructure realizations to accelerate solution of the Boundary Value Problems (BVP). Namely, we consider each tile to be a macro-element whose stiffness matrix is obtained using static condensation of internal unknowns of the finite-element (FE) stiffness matrix of the tile. Thus, the tile stiffness matrix is factorized only once at the beginning of the RVE size analysis. In the spirit of the Schur complement method, BVP of each microstructure realization then corresponds to a coarse grid problem composed of the macro-elements~\cite{kruis_domain_2006}, resulting in significantly less unknowns. In particular, the number of unknowns was reduced from 46 millions to 250 thousands for the largest investigated system in \Sref{sec:results}. Macro-elements also improve spectral properties of the final algebraic system, which is especially significant when investigating composites with high contrast in material properties of individual components. As a result, iterative solvers---such as the preconditioned conjugate gradient method used in this work---for the coarse grid problem require less iterations to achieve desired accuracy. 


Moreover, components of the apparent tensors are obtained as averaged dual quantities after solving BVP with a prescribed macroscopic tensor $\tens{\Strain}$ or $\tens{\Tempgrad}$, respectively, keeping one component equal to unity while the others remaining zero, recall~\Eref{eq:mechanical_def_stiffness}.
We also accelerate the averaging by constructing matrices that relate tile boundary degrees of freedom to the sought averages in the off-line phase. 

\section{Numerical tests}
\label{sec:results}

%
Performance of the proposed methodology and the sensitivity of the RVE size with respect to an investigated physical phenomenon, morphology of the microstructure, and contrast in constituent properties are illustrated with three distinct two-phase microstructures: a suspension of non-penetrable elliptic inclusions,~\Fref{fig:elliptic_inclusions}; a foam-like microstructure,~\Fref{fig:foam_microstructure}; and sandstone,~\Fref{fig:sandstone}. 

%
In order to circumvent the need for meshing complex geometries while maintaining the mesh compatibility across the relevant tile edges, we resorted to regular pixel-like grids. Each pixel represented a quadrilateral FE element with bilinear Lagrange basis functions. Following a sensitivity analysis of the tile apparent properties with respect to the mesh density, the resolution of each tile was determined as a compromise between accuracy and computational cost, see~\Tref{tab:resolution} for the chosen values. Because the first-order apparent properties are length-scale free,
%
%
we set the pixel size to be the reciprocal value of a tile resolution, resulting in a unitary tile size.

\begin{table}[!h]
	\caption{Single tile resolution and characteristics of the compressed microstructure systems.\replace{}{ Standard deviation of the volume fraction $\volfrac$ computed over the tile set is reported in parentheses.}}
	\centering
	\begin{tabular}{l|ccc}
		 & ellipse & foam & sandstone \\ \hline\hline
		resolution & 100$\times$100 px & 402$\times$402 px & 354$\times$354 px \\ 
		$\charlen$ & 0.157 & 0.047 & 0.051 \\
		$\volfrac$ & 0.359 \replace{}{(0.023)} & 0.742 \replace{}{(0.008)} & 0.169 \replace{}{(0.013)} \\
	\end{tabular}
	\label{tab:resolution}
\end{table}

%
\begin{figure*}[!ht]
	\centering
	\renewcommand{\arraystretch}{1.0}
	\setlength{\tabcolsep}{2pt}
	\begin{tabular}{ccc}
		\includegraphics[width=0.325\textwidth]{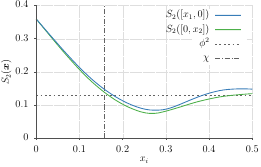} & \includegraphics[width=0.325\textwidth]{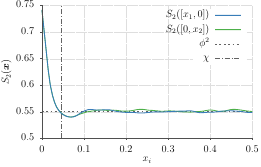} & \includegraphics[width=0.325\textwidth]{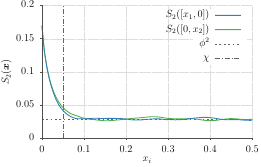} \\ 
		(a) & (b) & (c)
	\end{tabular}
	\caption{Sections of the inclusion two-point probability function $\Stwo$ along $x_1$ and $x_2$ axes and the highlighted characteristic length $\charlen$ of: (a)~the microstructure with mono-disperse elliptic inclusions, (b)~foam, and (c)~sandstone. In legends, $\volfrac$ denotes the volume fraction of the inclusions. Data was obtained by averaging $\Stwo$ statistics computed for 10 realizations of $40\times40$ tilings.}
	\label{fig:S2_and_char_length}
\end{figure*}
In order to relate the tile and RVE sizes to an intrinsic scale of a material, the characteristic length
\begin{linenomath}
\begin{equation}
	\charlen = \sqrt{ \frac{1}{\volfrac-\volfrac^2} \int_{\mathcal{T}} \lvert \Stwo(\tens{x}) - \volfrac^2 \rvert \de{\tens{x}} } \,,
	\label{eq:char_length}
\end{equation}
\end{linenomath}
was identified for each microstructure, similarly to~\cite{kanit_apparent_2006}.
In~\Eref{eq:char_length}, $\volfrac$ denotes the volume fraction of the inclusion phase\footnote{{We consistently refer to the continuous phase as a matrix and to the discontinuous phase as inclusions. Consequently, the volume fraction reported here for the foam microstructure is complementary to the standard notion of foam volume fraction.}} and $\Stwo\atx$ stands for the two-point probability function~\cite{torquato_random_2002}, which gives the probability of finding two points separated by $\tens{x}$ in the same constituent---the inclusion phase in our case. The integral in~\Eref{eq:char_length} is computed over $\mathcal{T} = \interval{0}{0.5} \times \interval{0}{0.5}$ in order to mitigate the effect of assembly-induced artefacts, see~\cite{doskar_aperiodic_2014} for further details. The characteristic lengths of the investigated microstructures, averaged from 10 realizations of $40\times40$ tilings, are plotted in~\Fref{fig:S2_and_char_length} and summarized in~\Tref{tab:resolution}.

%
Within our numerical tests, microstructure constituents were assumed isotropic. For the thermal conduction problem, the conductivity tensor $\tenss{K}$ of the $i$-th constituent then takes the form 
\begin{linenomath}
\begin{equation}
	\tenss{K}_i = \lambda_i \tenss{I}\,,
\end{equation}
\end{linenomath}
with $\lambda_i$ being the conductivity of the $i$-th phase and $\tenss{I}$ standing for the second order unit tensor. In the case of linear elasticity, the material stiffness tensor $\tensf{D}$ is given as
\begin{linenomath}
\begin{equation}
	\tensf{D}_i = \lambda_i \tenss{I} \otimes \tenss{I} + 2 \mu_i \tensf{I}^{s}\,,
\end{equation}
\end{linenomath}
where $\lambda_i$ and $\mu_i$ are the first and second Lamé coefficients of the $i$th phase, respectively, and $\tensf{I}^{s}$ denotes the fourth order unit tensor with major and minor symmetries. We further assumed plane strain conditions.

For each type of microstructure and each phenomenon, we investigated four material property contrasts $\contrast$ defined as
\begin{linenomath}
\begin{equation}
 \contrast = \frac{\lambda_2}{\lambda_1}\,.
\end{equation}
\end{linenomath}
The material parameters of the matrix-like phase (denoted with index 1 and depicted in dark gray colour in Figs.~\ref{fig:elliptic_inclusions}, \ref{fig:foam_microstructure}, and \ref{fig:sandstone}) were kept fixed at unity while the parameters of the second inclusion-like phase (indexed with 2 and shown in light gray colour) were proportionally scaled by the factors $0.01$, $0.1$, $10$, and $100$; see \Tref{tab:parameters}.

\begin{table}[!h]
	\caption{Combination of material parameters for individual contrasts}
	\centering
	\begin{tabular}{l|cc|cccc}
		$\contrast$ & \multicolumn{2}{c|}{thermal conductivity} & \multicolumn{4}{c}{linear elasticity} \\\hline
		& $\lambda_1$ & $\lambda_2$ & $\lambda_1$ & $\mu_1$ & $\lambda_2$ & $\mu_2$ \\ \hline\hline
		1:100	& 1 & 0.01 & 1 & 0.5 & 0.01 & 0.005 \\
		1:50	& 1 & 0.02 & 1 & 0.5 & 0.02 & 0.010 \\
		50:1	& 1 & 50 & 1 & 0.5 & 50 & 25 \\
		100:1	& 1 & 100 & 1 & 0.5 & 100 & 50 \\ 
	\end{tabular}
	\label{tab:parameters}
\end{table}

%
Finally, for scalar characterization of an apparent tensor, recall \Eref{eq:scalar_quantification}, we used the operator norm of the corresponding matrix representation, employing the Mandel notation in the case of linear elasticity. The proximity error $\errorProximity$ was calculated using the Frobenius norm in \Eref{eq:proximity_error}. Both significance levels $\confLevelOne$ and $\confLevelTwo$ were set to $0.01$ and the related limit errors were defined as $\errorSizeLim=0.01$ and $\errorProximityLim=0.05 \lVert \tenss{I} \rVert_{\textrm{Fro}}$, respectively, where the norm of a unitary tensor was used to cover consistently both thermal conductivity and linear elasticity. In all cases, we set $\realmin=5$; the upper limit was $\realmax=30$ for the first and third microstructure and $\realmax=25$ for the foam microstructure. Data reported in this Section follows from a single run of the proposed methodology. However, results of multiple runs with different random realizations (not reported here) show that the identified RVE size is consistent throughout different runs, albeit the number of realizations of intermediate SVE sizes may vary to accommodate the required accuracy $\errorSizeLim$. \replace{}{Especially for small SVEs, the scatter in the number of realization can be significant due to the random sampling.}

%
%

\subsection{Impenetrable elliptic inclusions}
\begin{figure}[ht]
	\centering
	\begin{tabular}{cc}
		\includegraphics[height=0.7\columnwidth]{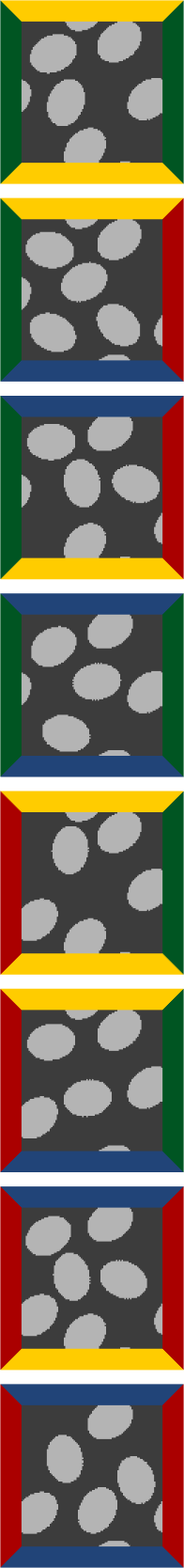} & \includegraphics[height=0.7\columnwidth]{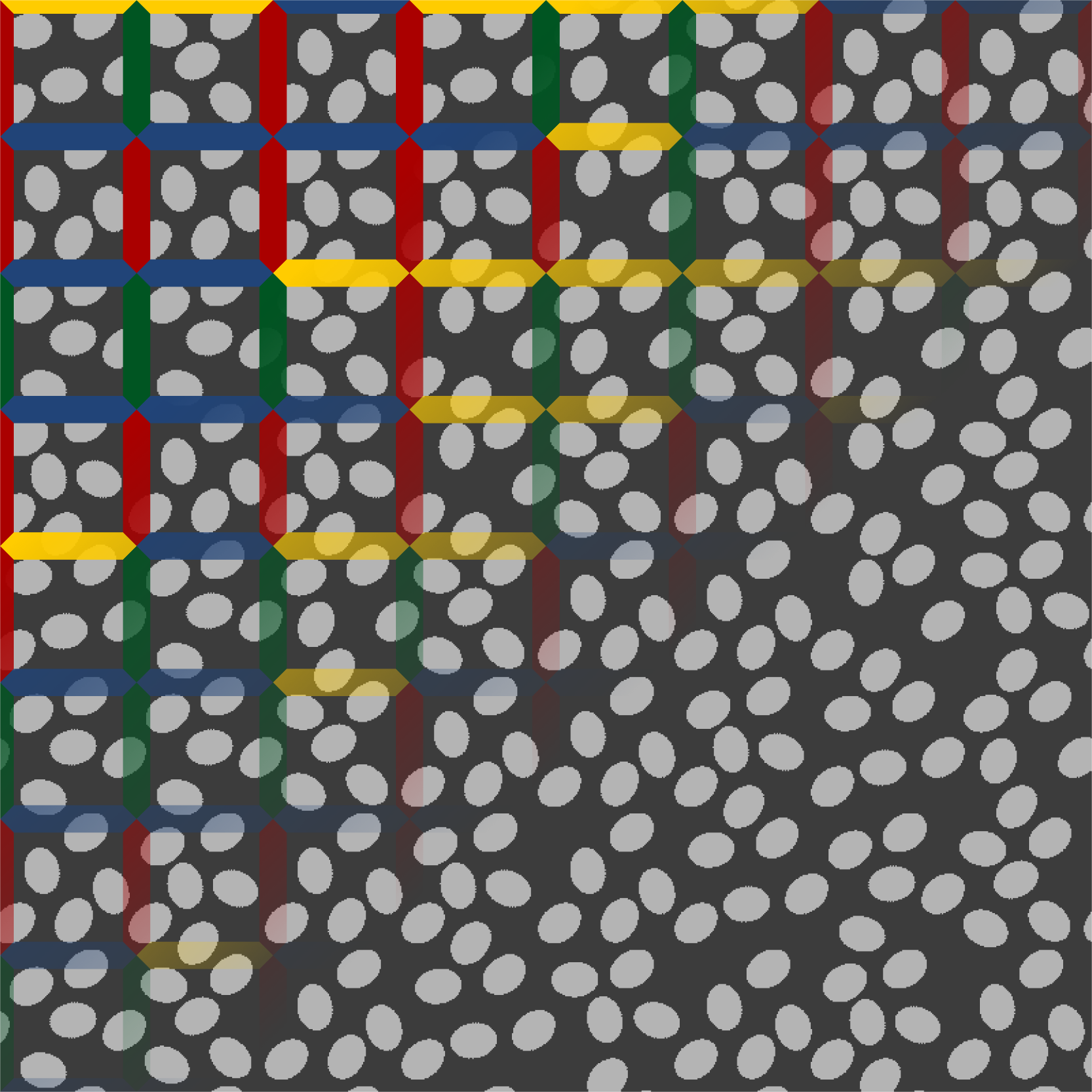} \\
		(a) & (b)
	\end{tabular}
	\caption{A microstructure with impenetrable mono-disperse elliptic inclusions compressed in a tiles set composed of 8 Wang tiles, (a), and a tiling sample with partially highlighted edge codes, (b).}
	\label{fig:elliptic_inclusions}
\end{figure}

First, we analysed a microstructure comprising impenetrable, mono-disperse elliptic inclusions of $0.75$ aspect ratio. The inclusion phase constituted $35.9~\%$ of the microstructure. Microstructural information was compressed in the set depicted in~\Fref{fig:elliptic_inclusions}, containing eight tiles with two edge codes on horizontal and vertical edges, respectively. 

Recall that the identified RVE size is always a multiple of the tile size, which defines the smallest attainable RVE size in turn. Thus, the tile-based approach is appealing particularly for problems with a high contrast $\contrast$ resulting in large RVEs. Here, the ratio $0.157$ between the characteristic length and the tile size, see~\Tref{tab:resolution}, allow us to investigate also 1:10 and 10:1 contrasts, which are neglected for the remaining two microstructures because these contrasts result in small RVE sizes of one or two tiles, dominated by the tile size rather than the RVE criterion.

\begin{figure}[!ht]
	\centering
	\renewcommand{\arraystretch}{1.0}
	\setlength{\tabcolsep}{0.5pt}
	\begin{tabular}{cc}
		\includegraphics[width=\halfColFigWidth]{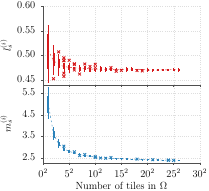} & \includegraphics[width=\halfColFigWidth]{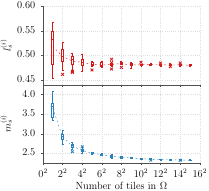} \\
		(a) & (b) \\
		\includegraphics[width=\halfColFigWidth]{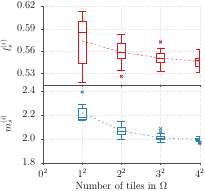} & \includegraphics[width=\halfColFigWidth]{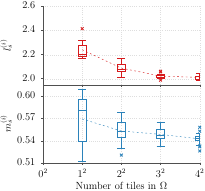} \\
		(c) & (d) \\
		\includegraphics[width=\halfColFigWidth]{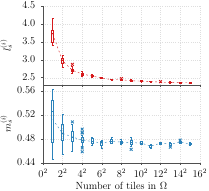} & \includegraphics[width=\halfColFigWidth]{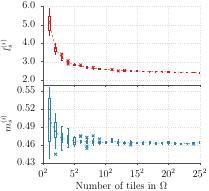} \\
		(e) & (f) \\
	\end{tabular}
	\caption{Box-and-whisker plots of the norms $\LscalOne$ and $\MscalOne$ for thermal conductivity of the microstructure with elliptic inclusions and contrasts in material properties: (a) 1:100, (b) 1:50, (c) 1:10, (d) 10:1, (e) 50:1, and (f)~100:1\,.}
	\label{fig:boxplot_E_TR}
\end{figure}

\begin{figure}[!ht]
	\centering
	\renewcommand{\arraystretch}{1.0}
	\setlength{\tabcolsep}{0.5pt}
	\begin{tabular}{cc}
		\includegraphics[width=\halfColFigWidth]{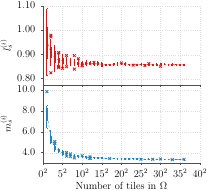} & \includegraphics[width=\halfColFigWidth]{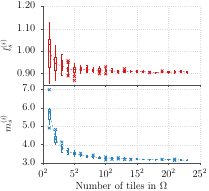} \\
		(a) & (b) \\
		\includegraphics[width=\halfColFigWidth]{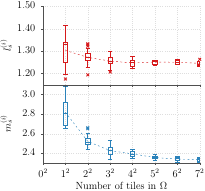} & \includegraphics[width=\halfColFigWidth]{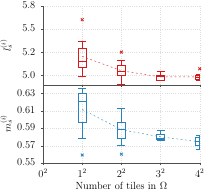} \\
		(c) & (d) \\
		\includegraphics[width=\halfColFigWidth]{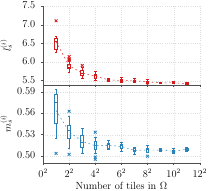} & \includegraphics[width=\halfColFigWidth]{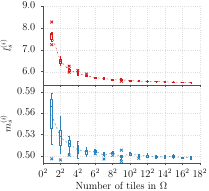} \\
		(e) & (f) \\
	\end{tabular}
	\caption{Box-and-whisker plots of the norms $\LscalOne$ and $\MscalOne$ for linear elasticity of the microstructure with elliptic inclusions and contrasts in material properties: (a) 1:0.01, (b) 1:0.05, (c) 1:0.1, (d) 10:1, (e) 50:1, and (f) 100:1\,.}
	\label{fig:boxplot_E_LE}
\end{figure}

The distribution of scalar quantities $\LscalOne$ and $\MscalOne$, characterizing the apparent properties of individual tiling realizations, are depicted using a box-and-whisker plot in~\Fref{fig:boxplot_E_TR} for the problem of thermal conductivity and in~\Fref{fig:boxplot_E_LE} for linear elasticity. Box boundaries and a mid-band denote the first and third quartile $Q_1$ and $Q_3$, and median $Q_2$, respectively; whisker ends mark an interval defined as $Q_2 \pm 1.5(Q_3-Q_1)$; and the crosses indicate potential data outliers. Dotted lines connect data averages. 

\begin{figure}[!ht]
	\centering
	\setlength{\tabcolsep}{0.5pt}
	\begin{tabular}{cc}
		\includegraphics[width=\halfColFigWidth]{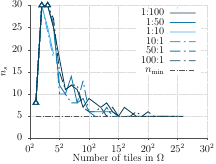} & \includegraphics[width=\halfColFigWidth]{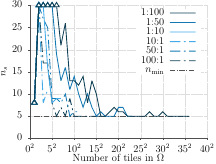} \\
		(a) & (b) \\
	\end{tabular}
	\caption{Number of SVE realizations for the microstructure with elliptic inclusions: (a) thermal conductivity, (b) linear elasticity.}
	\label{fig:realization_number_E}
\end{figure}

\begin{figure}[!ht]
	\centering
	\renewcommand{\tabcolsep}{0.5pt} 
	\begin{tabular}{cc}
		\includegraphics[width=\halfColFigWidth]{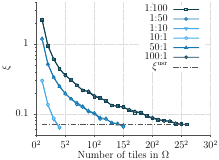} & \includegraphics[width=\halfColFigWidth]{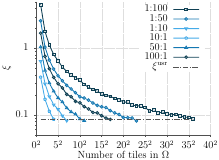}\\
		(a) & (b)
	\end{tabular}
	\caption{Convergence of the proximity error $\errorProximity$ with increasing tiling sizes for the microstructure with elliptic inclusions: (a) thermal conductivity, (b) linear elasticity.}
	\label{fig:convergence_E}
\end{figure}

The number of realizations of each SVE size required to meet the $\errorSizeLim$ criterion is given in~\Fref{fig:realization_number_E}; cases when the number of realizations was restricted by the upper limit $\realmax$ are denoted with empty triangle markers.
\replace{}{Note that due to the combinatorial nature of the SVE synthesis, the number of unique SVE realizations of given size $\svesize$ is limited. However, the number grows exponentially}\footnote{\replace{}{By construction, the tile sets in \mbox{\Fref{fig:elliptic_inclusions}(a)} and \mbox{\Fref{fig:foam_microstructure}(a)} allow for at least two distinct tiles for each position in a tiling; thus the number of unique SVEs is bounded from below by $2^{\svesize^2}$. For sandstone compression, see~\mbox{\Fref{fig:sandstone}(a)}, the lower bound is $4^{\svesize^2}$.}}\replace{}{ and therefore poses no practical restriction for larger SVEs.}
\replace{Also note that the}{The} number of $1\times 1$ realizations was limited by the number of individual tiles and the results are shown only for completeness. Finally, \Fref{fig:convergence_E} shows the convergence of the proximity error $\errorProximity$ to its limit value $\errorProximityLim$.

%
%

%
As expected, higher contrast in constituent properties led to increase in the RVE size, which ranged from four times the tile edge length for 1:10 and 10:1 contrasts up to 36 in the case of linear elasticity and contrast 100:1. The maximum number of realizations was set to 30 for this microstructure, which did not influence identification of the RVE size, see~\Fref{fig:realization_number_E}. 

In the particular problem of thermal conductivity, values $\LscalOne$ and $\MscalOne$ have direct physical interpretation as the largest and the inverse of the least principal conductivity. Comparing the values obtained for the RVE size indicates anisotropy in the homogenized material behaviour, which can be anticipated considering different $x_1$ and $x_2$ cross-sections of the two-point probability functions shown in~\Fref{fig:S2_and_char_length}a.

\subsection{Foam}
\begin{figure}[!ht]
	\centering
	\begin{tabular}{cc}
		\includegraphics[height=0.7\columnwidth]{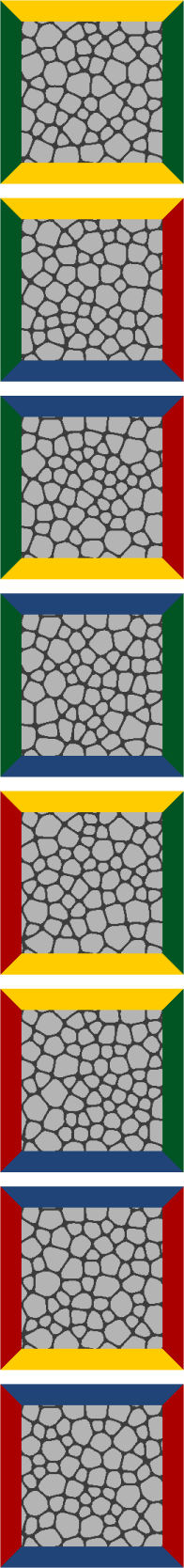} & \includegraphics[height=0.7\columnwidth]{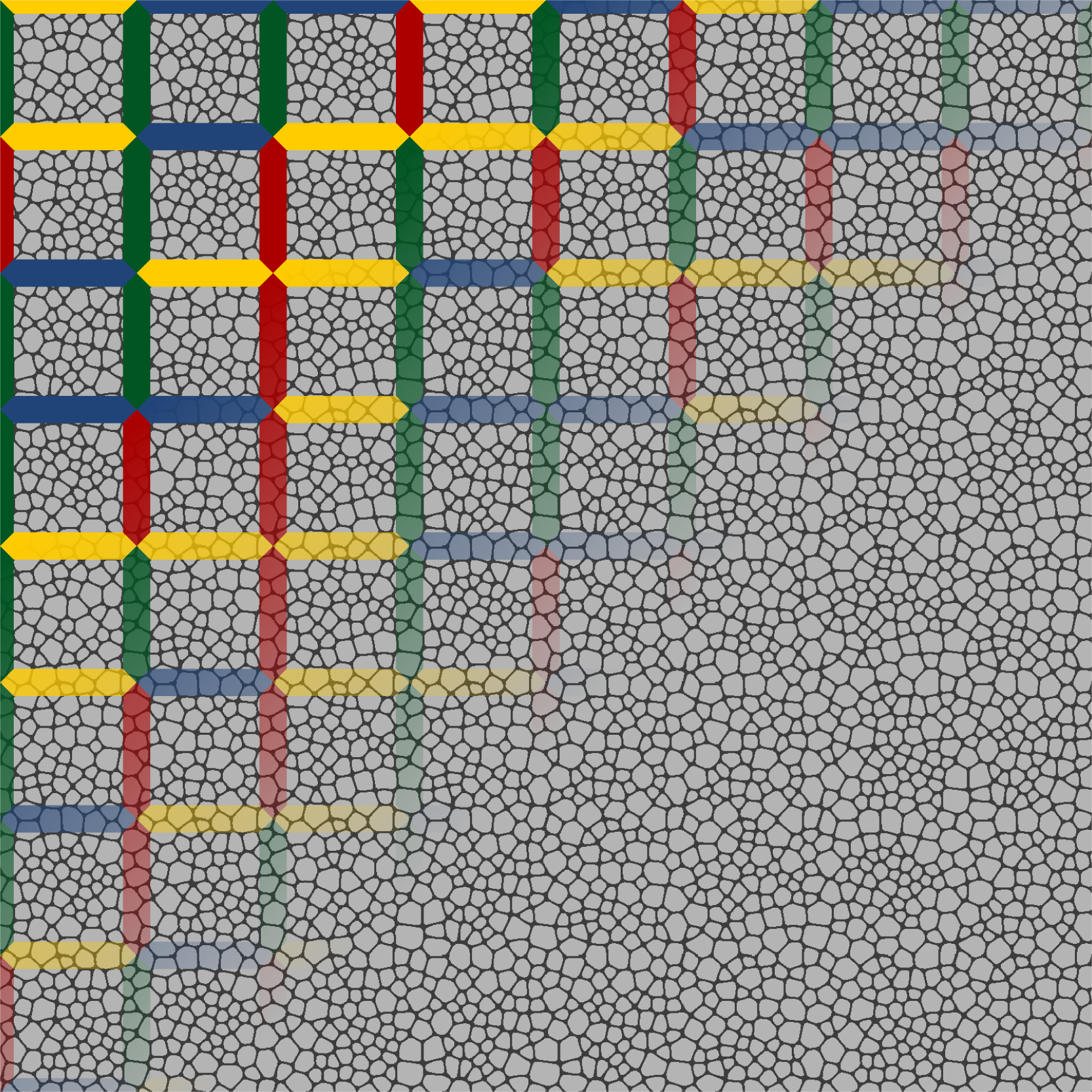} \\
		(a) & (b)
	\end{tabular}
	\caption{A foam-like microstructure compressed in a tiles set composed of 8 Wang tiles, (a), and a tiling sample with partially highlighted edge codes, (b).}
	\label{fig:foam_microstructure}
\end{figure}
\begin{figure}[!ht]
	\centering
	\renewcommand{\arraystretch}{1.0}
	\setlength{\tabcolsep}{0.5pt}
	\begin{tabular}{cc}
		\includegraphics[width=\halfColFigWidth]{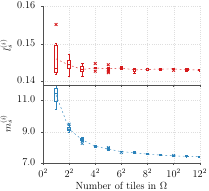} & \includegraphics[width=\halfColFigWidth]{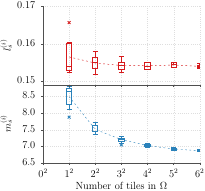} \\
		(a) & (b) \\
		\includegraphics[width=\halfColFigWidth]{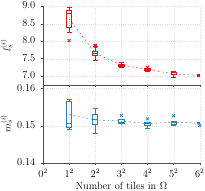} & \includegraphics[width=\halfColFigWidth]{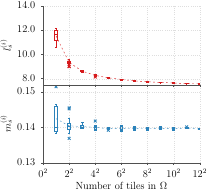} \\
		(c) & (d) \\
	\end{tabular}
	\caption{Box-and-whisker plots of the norms $\LscalOne$ and $\MscalOne$ for thermal conductivity of the foam-like microstructure and contrasts in material properties: (a)~1:100, (b)~1:50, (c)~50:1, and (d)~100:1\,.}
	\label{fig:boxplot_F_TR}
\end{figure}
\begin{figure}[!ht]
	\centering
	\renewcommand{\arraystretch}{1.0}
	\setlength{\tabcolsep}{0.5pt}
	\begin{tabular}{cc}
		\includegraphics[width=\halfColFigWidth]{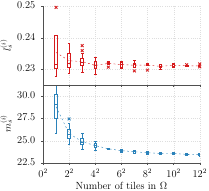} & \includegraphics[width=\halfColFigWidth]{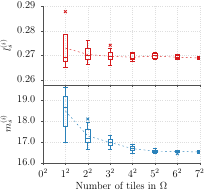} \\
		(a) & (b) \\
		\includegraphics[width=\halfColFigWidth]{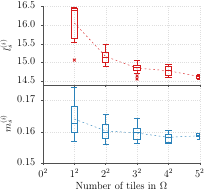} & \includegraphics[width=\halfColFigWidth]{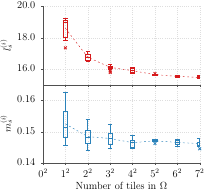} \\
		(c) & (d) \\
	\end{tabular}
	\caption{Box-and-whisker plots of the norms $\LscalOne$ and $\MscalOne$ for linear elasticity of the foam-like microstructure and contrasts in material properties: (a)~1:100, (b)~1:50, (c)~50:1, and (d)~100:1\,.}
	\label{fig:boxplot_F_LE}
\end{figure}
\begin{figure}[!ht]
	\centering
	\setlength{\tabcolsep}{0.5pt}
	\begin{tabular}{cc}
		\includegraphics[width=\halfColFigWidth]{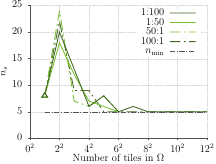} & \includegraphics[width=\halfColFigWidth]{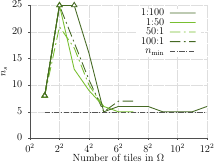} \\
		(a) & (b) \\
	\end{tabular}
	\caption{Number of SVE realizations for the foam-like microstructure: (a) thermal conductivity, (b) linear elasticity.}
	\label{fig:realization_number_F}
\end{figure}
\begin{figure}[!ht]
	\centering
	\setlength{\tabcolsep}{0.5pt}  
	\begin{tabular}{cc}
		\includegraphics[width=\halfColFigWidth]{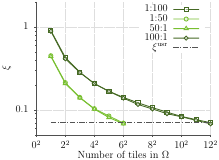} & \includegraphics[width=\halfColFigWidth]{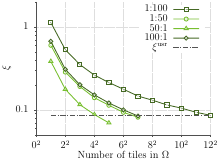} \\
		(a) & (b)
	\end{tabular}
	\caption{Convergence of the proximity error $\errorProximity$ with increasing tiling sizes for the foam-like microstructure: (a) thermal conductivity, (b) linear elasticity.}
	\label{fig:convergence_F}
\end{figure}
Motivated by our earlier study on elastic properties of aluminium foams~\cite{doskar_jigsaw_2016}, the second investigated microstructure was chosen to represent a two-dimensional sample of a closed cell foam. The system was compressed into the same tile set, in terms of tile code definition, as the previous microstructure. The internal geometry of tiles was artificially designed with a modified version of the level-set based approach developed by \citet{sonon_advanced_2015}. The compressed geometry, displayed in~\Fref{fig:foam_microstructure}, features large irregular inclusions separated with thin ligaments that form $25.8~\%$ of the microstructure volume. 

With infinite contrast, foams are typical representatives of complex materials with pronounced influence of actual microstructure composition on their overall response.
Our previous work~\cite{doskar_jigsaw_2016} corroborated this claim in two-dimensional setting, reporting the RVE size approximately in the order of magnitude of thousands of inclusion (voids) diameters. Even thought we do not assume the extreme case of voids, large RVE sizes are expected for higher contrasts as well.

Statistics of $\LscalOne$ and $\MscalOne$ for increasing tiling sizes are plotted in Figs.~\ref{fig:boxplot_F_TR}~and~\ref{fig:boxplot_F_LE} for thermal conductivity and linear elasticity, respectively. Number of realizations for each tiling size is shown in~\Fref{fig:realization_number_F} and the convergence towards the prescribed proximity limit $\errorProximityLim$ is depicted in~\Fref{fig:convergence_F}.
The identified RVE sizes ranged from three to 12 times the tile size in all analyses, the number of equi-sized realizations was limited to 25, see~\Fref{fig:realization_number_F}. Compared to the microstructure with elliptic inclusions, the nominal RVE sizes are smaller, however, the tile size is a characteristics of microstructure representation not its geometry. RVE sizes of different microstructures have to be compared to their characteristic length, see~\Fref{fig:scaled_convergence}, which supports the assumption of larger RVE than in the case of the microstructure with inclusions; see also additional discussion in Section~\ref{sec:results_discussion}. Finally, the converged values of $\LscalOne$ and $\MscalOne$ in~\Fref{fig:boxplot_F_TR} imply isotropy of the microstructure, observable also in~\Fref{fig:S2_and_char_length}b.

\subsection{Sandstone}
\begin{figure}[!ht]
	\centering
	\begin{tabular}{cc}
		\includegraphics[height=0.7\columnwidth]{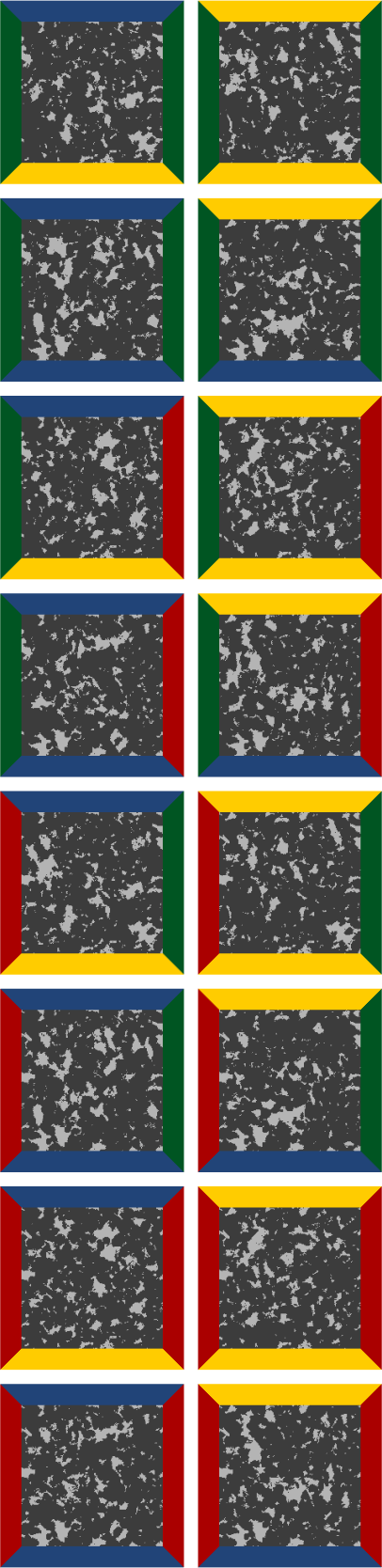} & \includegraphics[height=0.7\columnwidth]{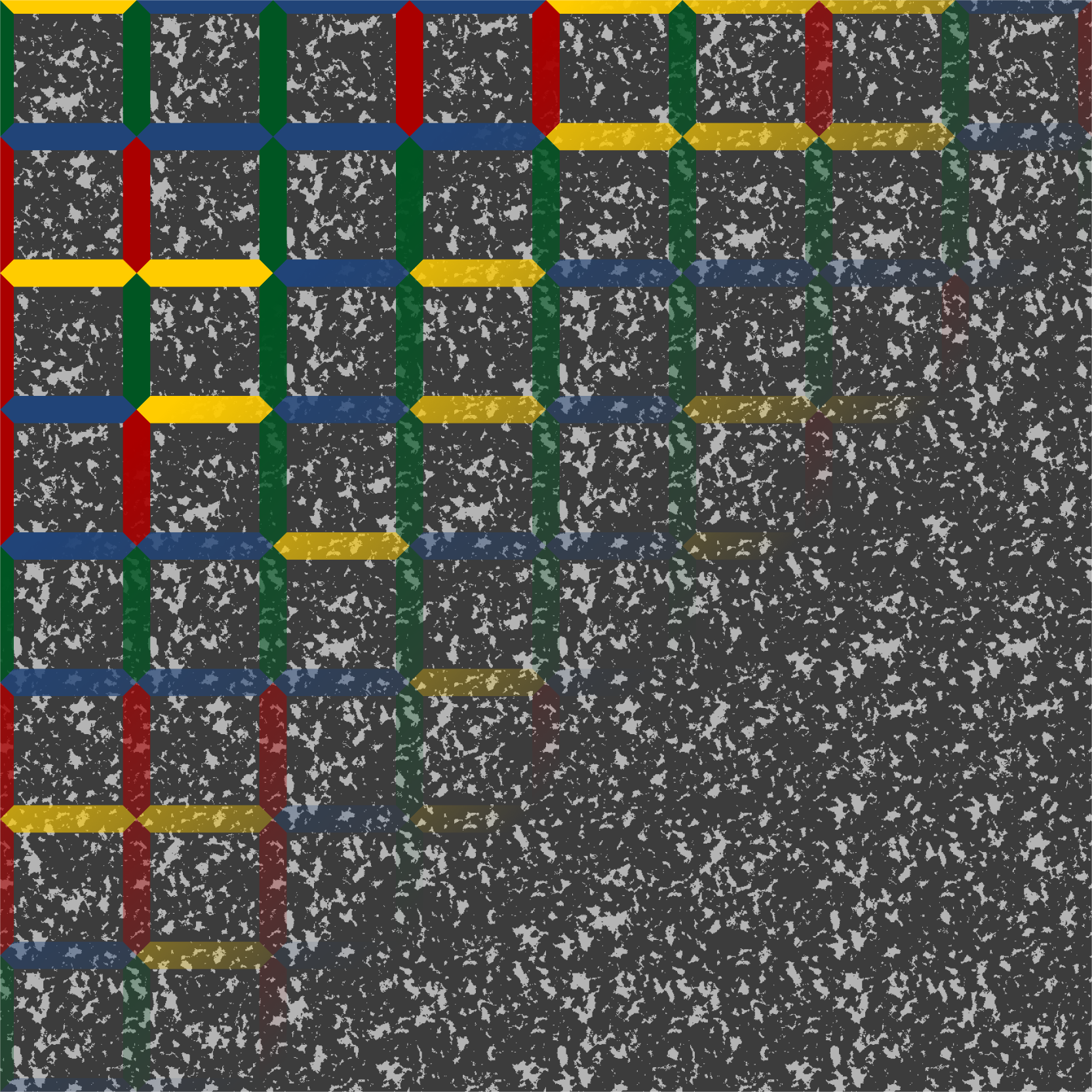} \\
		(a) & (b)
	\end{tabular}
	\caption{A microstructure of Gosford sandstone~\cite{sufian_microstructural_2013} compressed within a set of 16 tiles, (a), and a tiling sample with partially highlighted edge codes, (b).}
	\label{fig:sandstone}
\end{figure}
The last investigated microstructure is a two-dimensional representation of the Gosford sandstone studied in~\cite{sufian_microstructural_2013}. The microstructure was compressed using the sample-based quilting algorithm~\cite{doskar_aperiodic_2014}. Two distinct codes were assumed at horizontal and vertical edges. Based on our previous study~\cite{doskar_aperiodic_2014} and to emphasize that the tile set can be arbitrarily large, we used a richer tile set that contained all possible combinations of the codes. Similarly to the previous section, the second phase (displayed in light grey in~\Fref{fig:sandstone}) was originally void; in our parametric analysis, we reused only the geometry and assumed a solid second phase with parameters from~\Tref{tab:parameters}.
\begin{figure}[!ht]
	\centering
	\renewcommand{\arraystretch}{1.0}
	\setlength{\tabcolsep}{0.5pt}
	\begin{tabular}{cc}
		\includegraphics[width=\halfColFigWidth]{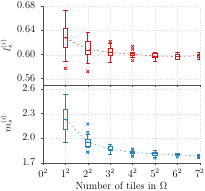} & \includegraphics[width=\halfColFigWidth]{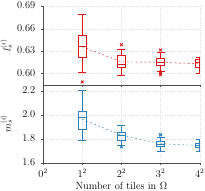} \\
		(a) & (b) \\
		\includegraphics[width=\halfColFigWidth]{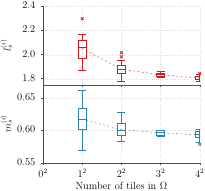} & \includegraphics[width=\halfColFigWidth]{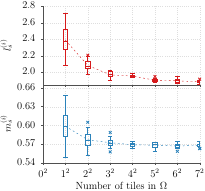} \\
		(c) & (d) \\
	\end{tabular}
	\caption{Box-and-whisker plots of the norms $\LscalOne$ and $\MscalOne$ for thermal conductivity of the sandstone microstructure and contrasts in material properties: (a)~1:100, (b)~1:50, (c)~50:1, and (d)~100:1\,.}
	\label{fig:boxplot_S_TR}
\end{figure}
\begin{figure}[!ht]
	\centering
	\renewcommand{\arraystretch}{1.0}
	\setlength{\tabcolsep}{0.5pt}
	\begin{tabular}{cc}
		\includegraphics[width=\halfColFigWidth]{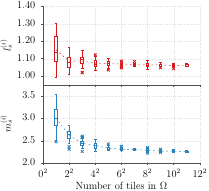} & \includegraphics[width=\halfColFigWidth]{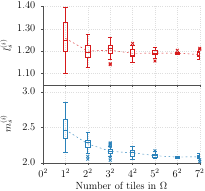} \\
		(a) & (b) \\
		\includegraphics[width=\halfColFigWidth]{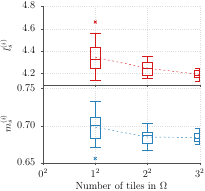} & \includegraphics[width=\halfColFigWidth]{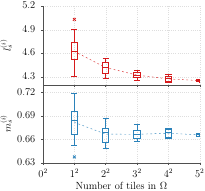} \\
		(c) & (d) \\
	\end{tabular}
	\caption{Box-and-whisker plots of the norms $\LscalOne$ and $\MscalOne$ for linear elasticity of the sandstone microstructure and contrasts in material properties: (a)~1:100, (b)~1:50, (c)~50:1, and (d)~100:1\,.}
	\label{fig:boxplot_S_LE}
\end{figure}
\begin{figure}[!ht]
	\centering
	\setlength{\tabcolsep}{0.5pt}
	\begin{tabular}{cc}
		\includegraphics[width=\halfColFigWidth]{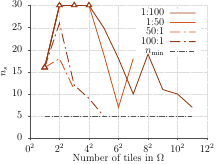} & \includegraphics[width=\halfColFigWidth]{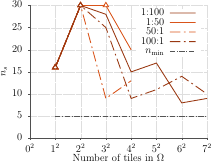} \\
		(a) & (b) \\
	\end{tabular}
	\caption{Number of SVE realizations for the sandstone microstructure: (a) thermal conductivity, (b) linear elasticity.}
	\label{fig:realization_number_S}
\end{figure}
\begin{figure}[!ht]
	\centering
	\setlength{\tabcolsep}{0.5pt}
	\begin{tabular}{cc}
		\includegraphics[width=\halfColFigWidth]{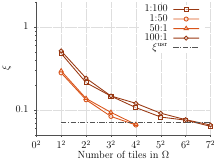} & \includegraphics[width=\halfColFigWidth]{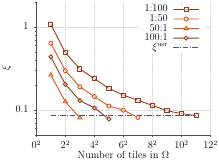} \\
		(a) & (b) \\
	\end{tabular}
	\caption{Convergence of the proximity error $\errorProximity$ with increasing tiling sizes for the sandstone microstructure: (a) thermal conductivity and (b) linear elasticity.}
	\label{fig:convergence_S}
\end{figure}

Again, scalar characterization of particular contrast settings and physical phenomena is summarized in Figs. \ref{fig:boxplot_S_TR} and \ref{fig:boxplot_S_LE}; the number of realizations generated for each tiling size is shown in~\Fref{fig:realization_number_S}; and \Fref{fig:convergence_S} depicts convergence of the RVE criterion. Low volume fraction of the second phase along with a large tile size to characteristic length ratio are likely the cause of small nominal RVE sizes, which were identified as seven times the tile size for contrasts 1:100 and 100:1 for thermal conductivity and twelve times the tile size in the case of linear elasticity and 1:100 contrast. On the other hand, more tiles in the compressed set and higher variability of the microstructure itself resulted in larger scatter of individual results, compare number of realizations in Figs. \ref{fig:realization_number_S} and \ref{fig:realization_number_F} and note that the characteristic length of the sandstone microstructure was similar to the foam-like microstructure. The converged values in~\Fref{fig:boxplot_S_TR} indicate slight anisotropy, which is emphasized when the inclusions are more conductive/stiffer than the matrix.

\subsection{Discussion}
\label{sec:results_discussion}
\begin{figure}[!ht]
	\centering
	\begin{tabular}{c}
		\includegraphics[width=\oneColFigWidth]{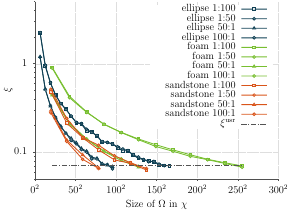}\\ (a)\\
		\includegraphics[width=\oneColFigWidth]{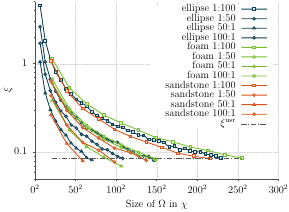}\\ (b)
	\end{tabular}
	\caption{Comparison of $\errorProximity$ convergence with increasing SVE sizes normalized against the characteristic length $\charlen$ for all investigated microstructures: (a) thermal conductivity, (b) linear elasticity.}
	\label{fig:scaled_convergence}
\end{figure}
\begin{figure}[!hb]
	\centering
	\includegraphics[width=\oneColFigWidth]{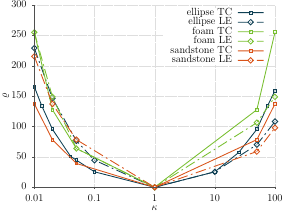}
	\caption{Lin-log plot of dependence of the scaled RVE-size $\RVEsizeCharLen$ on the contrast $\contrast$ of constituent properties for thermal conductivity (TC) and linear elasticity (LE).}
	\label{fig:RVE-contrast_dependence}
\end{figure}
%
%
As stated above, the nominal sizes of RVE identified in terms of tile multiples have to be scaled with the corresponding characteristic length $\charlen$ of the microstructure to allow mutual comparison. The scaled RVE sizes $\RVEsizeCharLen$ are plotted against the contrast $\contrast$ of constituent properties in~\Fref{fig:RVE-contrast_dependence}. Similarly, \Fref{fig:scaled_convergence} shows the $\errorProximity$ convergence lines from Figs.~\ref{fig:convergence_E}, \ref{fig:convergence_F}, and \ref{fig:convergence_S} as functions of the scaled size of~$\domain$.

%
Data in~\Fref{fig:scaled_convergence} confirms that the RVE size is indeed problem dependent and there are no universal scaling parameters common to both thermal conductivity and linear elasticity. For thermal conductivity, the inverse contrasts, i.e., 1:$n$ and $n$:1, resulted in nearly identical RVE sizes. On the other hand, for linear elasticity, cases with matrix stiffer than inclusions required approximately twice the RVE size, compared to the RVE size of the inverse contrast, to satisfy $\errorProximityLim$, compare Figs.~\ref{fig:scaled_convergence}a and \ref{fig:scaled_convergence}b or note the inclination of dash-dotted lines in~\Fref{fig:RVE-contrast_dependence}. Asymmetry was observed in a whole range of $\contrast$ and for all investigated microstructures.

%
\begin{figure}[!ht]
	\centering
	\includegraphics[width=\oneColFigWidth]{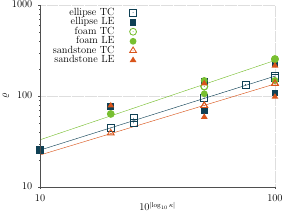}
	\caption{Log-log plot of the RVE-size dependence on the absolute contrast of constituent properties, defined as $10^{\lvert\log_{10}{\contrast}\rvert}$. The fit (solid lines) of the power-law relation (\ref{eq:RVE_fit}) is shown only for the thermal conductivity problem.}
	\label{fig:RVE-contrast_dependence_loglog}
\end{figure}
Plotting data from \Fref{fig:RVE-contrast_dependence} in a log-log graph~\Fref{fig:RVE-contrast_dependence_loglog} and modifying the horizontal axis such that the contrasts 1:$n$ and $n$:1 coincide reveal a power-law relation in the form
\begin{linenomath}
\begin{equation}
\RVEsizeCharLen = a \left(\contrast\right)^{b}\,,
\label{eq:RVE_fit}
\end{equation}
\end{linenomath}
with fitting parameters $a$ and $b$. The observed symmetry of the RVE size for thermal conductivity allows for replacing $\contrast$ with $10^{\lvert\log_{10}{\contrast}\rvert}$ in~\Eref{eq:RVE_fit} and consequently fitting only one set of parameters for the whole contrast range; linear elasticity problem requires separate handling of $\contrast<1$ and $\contrast>1$. 
\begin{table*}
	\caption{Values of Linear Least Square fit parameters from~\Eref{eq:RVE_fit}}
	\centering
	\begin{tabular}{r|ccc|ccc|ccc}
		& \multicolumn{3}{c|}{Ellipse} & \multicolumn{3}{c|}{Foam} & \multicolumn{3}{c}{Sandstone} \\
		& TC & LE ($\contrast<1$) & LE ($\contrast>1$) & TC & LE ($\contrast<1$) & LE ($\contrast>1$) & TC & LE ($\contrast<1$) & LE ($\contrast>1$) \\\hline
		a & 3.944 & 8.874 & 5.995 & 4.465 & 4.867 & 15.928 & 3.7457 & 11.910 & 3.292 \\
		b & 0.812 & -0.711 & 0.628 & 0.872 & -0.865 & 0.485 & 0.781 & -0.628 & 0.737 \\
		\# data & 11 & 4 & 3 & 5 & 3 & 2 & 5 & 3 & 2 \\
	\end{tabular}
	\label{tab:fitted_params}
\end{table*}
Parameters reported in~\Tref{tab:fitted_params} were obtained with a linear least square regression, taking into account also unreported results obtained for contrast 1:20 for all six combinations and 1:75, 1:25, 25:1, and 75:1 for the problem of thermal conductivity of the microstructure with elliptic inclusions. For the sake of brevity, the predicted fits are plotted in~\Fref{fig:RVE-contrast_dependence_loglog} only for the thermal conductivity problem. While the parameters $a$ and $b$ for thermal problems were obtained from ten and five data points, respectively, in certain cases fits for linear elasticity were based only on two data points and, thus, the identified values are inconclusive.

%
For a given contrast $\contrast$, the corresponding RVE size $\RVEsizeCharLen$ is smallest for the sandstone microstructure, followed by the microstructure with elliptic inclusions and the foam-like microstructure
. The ordering closely follows the volume fraction of the second phase; observe that the RVE sizes of the sandstone microstructure and the microstructure with elliptic inclusions are always closer together, compared to the foam-like microstructure. On the other hand, the influence of particular microstructure composition varies depending on a problem and chosen $\contrast$.

For the extreme contrasts and thermal conductivity, $\RVEsizeCharLen$ seems to be governed primarily by the volume fraction, due to the large difference between individual microstructures, and insensitivity to swapping material properties of the phases, see~\Fref{fig:scaled_convergence}a. On the contrary, in linear elasticity, complexity and the actual microstructure composition have a pronounced effect especially when the inclusions are stiffer; instances in which the matrix phase is stiffer seem to be dominated mainly by the contrast itself. 
%

%
%
\begin{figure}[!ht]
	\centering
	\begin{tabular}{c}		
		\includegraphics[width=\oneColFigWidth]{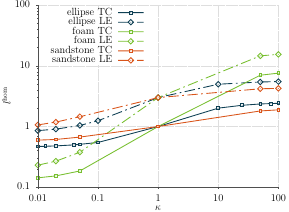}\\
		(a)\\
		\includegraphics[width=\oneColFigWidth]{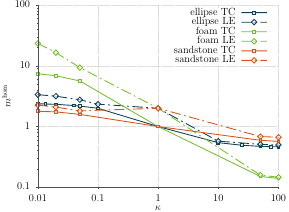}\\
		(b)
	\end{tabular}
	\caption{Log-log dependence of the norms $\LscalHom$ and $\MscalHom$ of the identified homogenized tensors, recall~\Eref{eq:statistical_form}, on contrast $\contrast$ of constituent properties. Values for unit contrast were computed from material characteristics of the matrix phase.}
	\label{fig:contrast-LMvalues_loglog}
\end{figure}
Influence of the inclusion parameters on the converged values $\LscalHom$ and $\MscalHom$, respectively, is shown in a log-log graph in~\Fref{fig:contrast-LMvalues_loglog}. Unlike the RVE-size dependence, curves in~\Fref{fig:contrast-LMvalues_loglog} indicate that no power law in the form of a monomial can be established except for the linear elasticity of the foam-like microstructure with $\contrast<1$. In all other cases, the overall scalar characteristics plateau soon and further increase of $\lvert\log_{10}(\contrast)\rvert$ does not lead to their significant change.

Also note in the box-and-whisker plots that the RVE size criterion $\errorProximity$ is driven by $\LscalMean$ for $\contrast>1$; and vice versa, the initial and final values of $\MscalMean$ differ more for $\contrast<1$.

%
Finally, the last observation regards the necessary number of realizations, shown in Figs.~\ref{fig:realization_number_E}, \ref{fig:realization_number_F}, and \ref{fig:realization_number_S}. Within a chosen microstructure, problems with similar final RVE sizes required approximately the same number of realizations at intermediate sizes. Moreover, as the SVE size approached the RVE one, the minimal number of realizations sufficed to meet $\errorSizeLim$ limit for all microstructures, except for sandstone. 

Thus, the RVE size, at least in the context of our methodology, seems to be driven mainly by the ensemble average and not its variation, which corroborates conclusions of Moussaddy\etal{}~\cite{moussaddy_assessment_2013}, who warned against using the variance of apparent properties as the only RVE criterion.

\section{Summary}
\label{sec:summary}

We demonstrated that the compressed representation of materials with random microstructure by means of Wang tiles is an appealing framework for numerical homogenization and problems of RVE size determination, in particular. Upon an off-line phase of compressing the microstructural information into a set of tiles, the framework facilitates instant on-line random generation of statistically coherent realizations of compressed microstructures. Moreover, adopting elemental ideas of domain decomposition and considering each tile as a macro-element reduce significantly the number of degrees of freedom and improve the condition number of the resulting algebraic system, a desirable feature especially when dealing with highly contrasted problems.

With the emphasis on obtaining bounds on the effective property, we established a methodology that identifies the RVE size for a user-defined accuracy. The methodology benefits from a regular partitioning inherent to the tiling concept and directly utilizes the Partition theorem and statistical sampling to construct confidence intervals of the apparent properties. The proposed methodology works at two levels:
\begin{enumerate}
	\item For a fixed domain size, new microstructure realizations are generated on-the-fly and their apparent properties are computed until the confidence intervals narrow below a user-defined threshold;
	\item The convergence criterion is checked and the algorithm either moves to larger domains or terminates identifying the sought RVE size. The methodology takes into account both the statistical deviation of apparent properties and the discrepancy between their mean values. This makes it robust against premature convergence.
\end{enumerate}

The efficiency of the Wang tiling concept allowed us to illustrate the methodology with a large set of problems. We performed the RVE size identification for three materials from a class of microstructures with clearly identifiable matrix- and inclusion-like phases, yet of different volume fraction and complexity of internal composition. For each material, we investigated two homogenization problems---linear heat conductivity and elasticity. Unlike the majority of similar studies, we kept the volume fraction of inclusions fixed and scaled the ratio between the matrix and inclusion material parameters. Without claiming observations to be general rules, for our particular setting, the RVE size seems to be driven mainly by the mean values of the apparent properties, which corroborates conclusions of~\citet{moussaddy_assessment_2013}. The effect of the actual microstructure composition and the role of matrix or inclusion material is significantly pronounced in the case of linear elasticity, while swapping the constituents has negligible effect on the RVE size for heat conductivity. Data also indicates a power-law relation between the RVE size expressed in terms of the microstructure characteristic length $\charlen$ and the contrast $\contrast$ of constituent properties. However, note that the power-law relation common to both the thermal and linear elasticity problems may stem from the adopted simultaneous scaling of Lam\'{e} coefficients in~\Tref{tab:parameters}, which resembles the scaling of the conductivity coefficient. Conversely, a scaling in the form of a simple monomial cannot be established for the converged scalar characteristics of the overall material behaviour but for the compliance of the foam-like microstructure with $\contrast<1$.

\replace{}{The present approach directly extends to linear three-dimensional problems, in which the acceleration through pre-computed factorization of each tile will be even more pronounced. However, robust methods for compressing three-dimensional microstructures, complemented with a tool for generating topologically and geometrically consistent discretization of Wang cubes, are yet unavailable and constitute our current work.}

\section*{Acknowledgement}

Martin Doškář and Jan Novák acknowledge the endowment of the Ministry of Industry and Trade of the Czech Republic under the project No. FV10202.
Jan Zeman was partially supported by the Czech Science Foundation, under the projects Nos. 17-04150J and 17-04301S.
We would also like to thank Ond\v{r}ej Roko\v{s} from TU Eindhoven and anonymous reviewers for valuable comments on the initial version of the manuscript.

\bibliography{bibliography}
\bibliographystyle{my-elsarticle-num-names}
\biboptions{square,numbers,sort&compress}

\end{document}